# A hybrid framework integrating classical computers and quantum annealers for optimisation of truss structures

Van-Dung Nguyen[a,b,*], Erin Kuci[a], Michel Rasquin[a], Ludovic Noels[b]

*[a]Centre de recherche en aéronautique (Cenaero) Rue des Frères Wright, 29, 6041 Gosselies, Belgium*

*[b]Computational & Multiscale Mechanical of Materials (CM3), Department of Aerospace and Mechanical Engineering, University of Liège, Quartier Polytech 1, Allée de la Découverte 9, B-4000 Liège, Belgium*

[*] *Corresponding author*
***Email addresses**:*
*vandung.nguyen@cenaero.be (Van-Dung Nguyen)*
*erin.kuci@cenaero.be (Erin Kuci)*
*michel.rasquin@cenaero.be (Michel Rasquin)*
*l.noels@ulg.ac.be (Ludovic Noels)*

## Abstract

This work proposes a hybrid framework combining classical computers with quantum annealers for structural optimisation. At each optimisation iteration of an iterative process, two minimisation problems are formulated, one for the underlying mechanical boundary value problem through the minimisation potential energy principle and one for the minimisation problem to update the design variables. Our hybrid approach leverages the strength of quantum computing to solve these two minimisation problems at each step, thanks to the developed quantum annealing-assisted sequential programming strategy introduced in [Nguyen, Wu, Remacle, and Noels. A quantum annealing-sequential quadratic programming assisted finite element simulation for non-linear and history-dependent mechanical problems. European Journal of Mechanics-A/Solids 105 (2024): 105254]. The applicability of the proposed framework is demonstrated through several case studies of truss optimisation, highlighting its capability to perform optimisation with quantum computers. The proposed framework offers a promising direction for future structural optimisation applications, particularly in scenarios where the quantum computer could resolve the size limitations of the classical computers due to problem complexities.

*Keywords:* Quantum computing, Quantum annealing, Optimisation, Truss structure

## Nomenclature

| | | | |
|---|---|---|---|
| $N, M$ | Number of bars and of nodes in the truss structure | $\boldsymbol{\lambda}$ | Auxiliary variables associated to the inequality constraints |
| $L_k, A_k, \alpha_k$ | Length, cross-sectional area, and cross-sectional ratio of the kth bar | $L$ | Number of qubits per continuous variable in the scalar continuous-binary approximation |
| $\boldsymbol{\alpha}, \boldsymbol{\alpha}_{\min}, \boldsymbol{\alpha}_{\max}$ | Vector of all cross-sectional area ratios -design variables- and its lower and upper bounds | $\sigma, \kappa, \zeta_{\blacksquare}$ | Parameters for general continuous-binary approximation |

| $V^{\text{target}}$ | Target material volume | $\bar{\delta}, \epsilon$ | Central value and discretization error of the scalar continuous-binary approximation |
|---|---|---|---|
| $\boldsymbol{K}, \boldsymbol{U}$ | Stiffness matrix and nodal displacement vector | $\bar{\boldsymbol{\delta}}, \boldsymbol{\epsilon}$ | Central value and discretization error of the vector continuous-binary approximation |
| $\Psi, \mathrm{C}$ | Potential energy and work of external force | $\boldsymbol{g}, \boldsymbol{A}$ | Gradient and Hessian scaled by a factor of 0.5 of the objective function |
| $\boldsymbol{H}_{\blacksquare}, \lvert\phi_{\blacksquare}\rangle$ | Hamiltonian and quantum state vector | $\boldsymbol{\epsilon}_0$ | Initial discretization error |
| $K$ | Number of qubits | $\xi$ | Shrinking factor |
| $A, B$ | Annealing schedule functions | $N_{\text{steps}}$ | Maximum number of iterations. |
| $\boldsymbol{H}_f^{\text{Ising}}$ | Ising Hamiltonian | $N_{\text{fails}}$ | Allowable number of failed iterations |
| $\boldsymbol{h}, \boldsymbol{J}$ | Ising Hamiltonian user-defined parameters | $s$ | Gradient of $C$ at $k^{th}$ iteration |
| $\boldsymbol{F}_{\text{Ising}}, \boldsymbol{s}$ | Ising function and vector of spin variables | $h$ | Material volume constraint |
| $\boldsymbol{F}_{\text{QUBO}}, \boldsymbol{b}, \boldsymbol{Q}$ | QUBO function and vector of binary variables, and QUBO matrix | $\boldsymbol{D}^{(k)}$ | Gradient of $h$ at $k^{th}$ iteration |
| $f, h_{\blacksquare}, l_{\blacksquare}$ | General multivariate functions used to denote objective functions, equality constraint, inequality constraint | $K^{\text{ME}}$ | Number of qubits used to solve the mechanical problem |
| $\boldsymbol{x}, \boldsymbol{x}_{\text{min}}, \boldsymbol{x}_{\text{max}}$ | Optimised argument and its lower and upper bounds | $K^{\text{DS}}$ | Number of qubits used to solve the design update problem |
| $\boldsymbol{\delta}, \boldsymbol{d}_{\text{min}}, \boldsymbol{d}_{\text{max}}, W$ | Incremental argument, its lower and upper possible values and its dimension | $T_{\text{QA}}^{\text{ME}}, T_{\text{QA}}^{\text{DS}}$ | Average time necessary to complete a mechanical step and a design update step using quantum annealing |
| $\boldsymbol{\delta}_{\text{min}}^{\text{b}}, \boldsymbol{\delta}_{\text{max}}^{\text{b}}$ | Lower and upper bounds defined the hyper-box | $T_{\text{SA}}^{\text{ME}}, T_{\text{SA}}^{\text{DS}}$ | Average time necessary to complete a mechanical step and a design update step using simulated annealing |
| $\Phi^{\blacksquare}(\blacksquare; \blacksquare), P$ | Polynomial approximation operator and its polynomial order | | |

## 1. Introduction

A truss structure is an architectural network of interconnected bars commonly found in civil and mechanical engineering, such as bridges, roofs, towers, etc. Truss optimisation aimed at finding the most efficient configuration of a truss structure under given loading conditions while minimising material usage and ensuring its strength, stability, and performance [1–4]. The optimisation process involves not only finding the best layout and connectivity of the truss members but also optimising the material distribution in each member. For truss optimisation performed with traditional computers, an increasing complexity of the problems requires extensive computational resources.

Recent advances in quantum computing hardware promise revolutionary solutions for solving engineering applications by leveraging quantum mechanics such as superposition and entanglement to perform computations [5, 6]. Unlike classical computers, where the information is stored and manipulated by bit-strings consisting of 0 and 1, quantum computing uses quantum bits (so-called qubits), which can exist in a superposition of 0 and 1 at the same time. As a result, $K$ qubits can represent $2^K$ bit-strings simultaneously, whilst $K$ classical bits can represent only one bit-string, leading to the exponential increase of the quantum computing power. Quantum computing offers the potential to revolutionise numerical computations that are currently beyond the hand of classical computing. There exist two main concepts for implementing quantum hardware: gate-based quantum computer and quantum annealer.

Gate-based quantum computers allow manipulating the qubits through a series of quantum logic gates, which are analogous to classical logic gates [7, 8]. They theoretically enable universal computations. An extensive number of quantum algorithms has been developed to run with fault tolerant systems, for instance, Shor's factorisation algorithm [9], Grover's quantum search algorithm [10], and Harrow-Hassidim-Lloyd (HLL) algorithm [11] for solving a system of linear equations. However, such fault-tolerant quantum devices are not available yet; the existing gate-based quantum computers are referred to as noisy-intermediate-scale quantum (NISQ) devices, for which variational quantum algorithms (VQA), such as variational quantum eigen solver [12] and quantum approximate optimisation algorithm (QAOA) [13], are suitable. In general, VQA is a hybrid approach combining quantum and classical computing to find parameters of parametrised quantum circuits with classical optimisers to minimise a cost function whose value is obtained upon quantum measurements on these circuits [14]. The applicability of VQAs in topological optimisation has been explored [15–17]. Quantum computing also promises to solve computational mechanics problems [18, 19]. Nevertheless, gate-based quantum computers are still in the early stages of development since the current quantum hardware is very sensitive to noise and surfers from high error rates, which makes it challenging to execute quantum algorithms accurately and efficiently.

Quantum annealer is particularly designed to solve binary quadratic models, which can be formulated as Ising or Quadratic Unconstrained Binary Optimization (QUBO) problems, by taking advantage of the quantum annealing (QA) technologies [20–23]. The qubits are initialised at a simple state, and their interactions or entanglements are progressively modified to reach the ground state of the encoded problem, which corresponds to the global minimisation state of the problem at hand. To reach this minimum state, QA considers the quantum tunnelling effect in which a particle has the ability to pass through a potential energy barrier, which could not classically be overcome due to insufficient energy, to explore the energy landscape in order to reach its ground state [6]. Compared to gate-based quantum computers, quantum annealers are more resilient against noise but much less versatile since they allow only considering unconstrained binary optimisation (Ising or QUBO) problems.

Quantum annealing devices were applied to various applications across multiple disciplines, see reviews by [24, 25]. The problems are however required to be transformed into Ising or QUBO forms to be embedded in the quantum annealers. For this purpose, the problem under consideration is first formulated as minimising an objective function of continuous and/or discrete variables and the latter is then transformed into QUBO. When the problem can be naturally expressed under a linear/quadratic form of the unknowns, QUBO can be achieved directly using a continuous-binary linear transformation, e.g. for problems such as linear regressions [26, 27] with quadratic cost function and linear elastic finite element simulations [28–30] with quadratic potential energy. In a general situation in which the function to be minimised is not quadratic, a second-order Taylor's series followed by a continuous-binary linear transformation to obtain a QUBO is iteratively carried out, leading to the so-called quantum annealing-assisted sequential quadratic programming (QA-SQP) framework [31]. QA-SQP is an iterative method, which offers the advantage of controlling the resolution error resulting from the annealing procedure. In this work, this advantage of the QA-SQP framework is exploited for the optimisation of truss structures.

Recently, building on its ability to find ground states, quantum annealing has been successfully applied to structural optimisation with proof of concepts. There have been few studies exploring this opportunity [32–37], in which the difference between them relies mainly on how to translate an objective function into single or a series of QUBO forms. Within a statistically admissible finite element formulation, Key and Freinberger [37] addressed the mechanical equilibrium and compliance minimization simultaneously by formulating a unique minimization problem defined in terms of both stresses and design variables. Wils & Chen [32] considered a symbolic finite element approach to express the objective function under a fractional form whose minimisation follows an iterative procedure of minimising adaptive non-fractional functions which are subsequently transformed into a QUBO form [38]. Ye et al. [33] combined the QA and the generalised Bender's decomposition method [39] to transform the original problem into a sequence of mixed-integer linear programming (MILP) problems, which are then reformulated as QUBO. Honda et al. [34] considered the combinatorial random number sums [40] to convert the objective functions to QUBO. Wang et al. [35] considered the direct mapping between the total mass of the structure to be minimised with a discrete set of cross-sectional areas to form QUBOs. A hybrid framework for both truss and continuum structures optimisation were proposed by [36] in which the structural analysis is performed on classical computers while the QA is used for topology updates. When performing the topology updates, the corresponding QUBO is derived by maximising the structural stiffness, leading to a linear form of the objective functions with respect to the design variables. These works provide substantial evidence of QA's potential in topology optimization. However, there remains a research gap in the literature for an efficient scheme that directly consider QA results not only to update the design structure, but also to solve the partial different equations governing the structure equilibrium state, which is the objective of this paper in the context of truss structures.

The QA-SQP framework [31] is extended in this paper and applied to the optimisation of truss structures. A Taylor's series is used to approximate the constrained minimisation problem at hand under a polynomial form, which becomes a quadratic binary form using a continuous-binary linear transformation. Such a quadratic form can be transformed directly into QUBO. For this purpose, the constraints are accounted for by quadratic penalty functions. Also, to guarantee the quality of this approximation, a hyper-box is defined to limit the approximation neighbourhood. The structure analysis is reformulated as a minimisation problem thanks to the principle of minimum potential energy [41], which is then solved with the QA-SQP framework. As a result, each optimisation step consists of solving two consecutive minimisation problems with QA-SQP, one for the structure analysis and one for updating the design variables.

The paper is organised as follows. Section 2 recalls the problem setting of the optimisation problem. The principle of quantum annealing is briefly summarised in Section 3. Section 4 unveils our QA-SQP framework and its exploitation in the optimisation problem. Several benchmarks for both two-dimensional and three-dimensional cases, and in the context of both linear and nonlinear material responses, are provided in Section 5 to demonstrate the capability of the proposed framework. Finally, some conclusions and perspectives are drawn in Section 6.

## 2. Optimisation problem setting

We consider a truss structure of $N$ bars interconnected at $M$ nodes in $R^d$ with $d = 2$ or 3. We denote $L_k$ and $A_k$ as the length and cross-sectional area of the $k^{th}$ member, respectively, with $k = 1, \ldots, N$. The cross-sectional area $A_k$ is parametrised by a scalar $\alpha_k$, such that

$$A_k = \alpha_k A_k^0, \tag{1}$$

where $A_k^0$ is the cross-sectional area of $k^{th}$ member in the initial design. We denote

$$\boldsymbol{\alpha} = [\alpha_1 \; \alpha_2 \ldots \alpha_N]^T, \tag{2}$$

as the vector of design variables.

This structure is subjected to an external nodal force vector $F \in \mathbb{R}^{dM}$ at its $M$ nodes. The structure balance results in the following equation

$$\boldsymbol{K}(\boldsymbol{\alpha})\boldsymbol{U} = \boldsymbol{F}, \tag{3}$$

where $\boldsymbol{K}$ is the global stiffness matrix being function of $\boldsymbol{\alpha}$ and $\boldsymbol{U}$ is the kinematically admissible nodal displacement vector. Following the parametrisation (1), the elementary stiffness matrix of the $k^{th}$ bar reads

$$\boldsymbol{K}_k^{\mathrm{el}} = \alpha_k \boldsymbol{K}_k^{\mathrm{el},0}, \tag{4}$$

where $k = 1, \cdots, N$ and $\boldsymbol{K}_k^{\mathrm{el},0}$ is its elementary stiffness matrix in the initial design. As a result, the global stiffness matrix $\boldsymbol{K}$ in Eq. (3) is achieved through the assembling process of its elementary counterparts as

$$\boldsymbol{K}(\boldsymbol{\alpha}) = \bigwedge_{k=1}^{N} \alpha_k \boldsymbol{K}_k^{\mathrm{el},0} = \sum_{k=1}^{N} \alpha_k \boldsymbol{K}_k \,, \tag{5}$$

where $\wedge$ is the assembling operator and $\boldsymbol{K}_k$ is the stiffness $\boldsymbol{K}_k^{\mathrm{el},0}$ assembled in the global system. Eq. (3) has the solution

$$\boldsymbol{U} = \boldsymbol{K}^{-1}\boldsymbol{F}, \tag{6}$$

which depends on the design variables α and generally cannot be expressed in an explicit form. Following the principle of minimum potential energy [41], Eq. (3) can be rewritten as minimising the potential energy following

$$\boldsymbol{U}(\boldsymbol{\alpha}) = \arg\min_{\boldsymbol{U}'} \Psi(\boldsymbol{U}'; \boldsymbol{\alpha}) \,, \tag{7}$$

where $\Psi(\boldsymbol{U}; \boldsymbol{\alpha}) = \frac{1}{2}\boldsymbol{U}^T\boldsymbol{K}\boldsymbol{U} - \boldsymbol{F}^T\boldsymbol{U}$ is the potential energy.

The optimisation problem is stated as finding the cross-section of each bar in the structure, which minimises the work of the external force defined by

$$C(\boldsymbol{\alpha}) = \boldsymbol{F}^T \boldsymbol{U}(\boldsymbol{\alpha}), \tag{8}$$

while respecting the volume constraint. This optimisation problem is formally stated as follows:

$$\boldsymbol{\alpha} = \arg\min_{\boldsymbol{\alpha}'} C(\boldsymbol{\alpha}'),$$
$$\text{subject to} \begin{cases} \text{Eq. (7)} \\ \sum_{k=1}^{N} \alpha_k L_k A_k^0 - V^{\text{target}} = 0 \\ \boldsymbol{\alpha}_{\min} \leq \boldsymbol{\alpha} \leq \boldsymbol{\alpha}_{\max} \end{cases} \tag{9}$$

where in this context, $\boldsymbol{U}$ is the so-called state variables to be distinguished with $\boldsymbol{\alpha}$ being the design variables. The first constraint is to ensure that the structure is in static equilibrium under given loading conditions. The second constraint enforces the volume of material to be equal to $V^{\text{target}}$. The last constraint specifies the lower-bounds $\boldsymbol{\alpha}_{\min}$ and upper bounds $\boldsymbol{\alpha}_{\max}$ of the design variable $\boldsymbol{\alpha}$.

## 3. Quantum annealing technique

Quantum annealer is a hardware implementation of the quantum annealing process. In quantum computing, the fundamental unit is the quantum bit (so-called qubit)[1]. Unlike classical bit which can take only one of the two the binary states $|0\rangle$ and $|1\rangle$ in classical computers, a qubit can represent $|0\rangle$, $|1\rangle$, and their superposition, *i.e.* $x|0\rangle + y\,|1\rangle$ in which $x$ and $y$ are complex numbers satisfying $|x|^2 + |y|^2 = 1$, and $|x|^2$ and $|y|^2$ are the probabilities to be in the state $|0\rangle$ and $|1\rangle$, respectively. This quantum characteristic allows for the simultaneous data processing.

The objective of quantum annealing is to solve the ground-state problem of a quantum-mechanical system characterised by the Hamiltonian Hf, which corresponds to the lowest energy eigenstate of $\boldsymbol{H}_f$, *i.e.*

$$|\phi_0\rangle = \arg\min_{|\phi\rangle} \langle\phi|\boldsymbol{H}_f|\phi\rangle, \tag{10}$$

where $\boldsymbol{H}_f$ is the Hamiltonian operator and $|\phi\rangle$ is a vector characterising the quantum state of the system. Starting from a simple ground state of the initial Hamiltonian $\boldsymbol{H}_i$ and gradually varying it according to a time-dependent Hamiltonian, the adiabatic theorem [42] ensures that the system remains close to the ground state throughout the entire evolution, provided the change is slow enough. Consequently, one can follow the following algorithm to find the ground state of the Hamiltonian $\boldsymbol{H}_f$ as follows:

- The system is initialised with the ground state of a simple Hamiltonian $\boldsymbol{H}_i$ whose known quantum ground state can be easily prepared.
- The system is slowly modified following an adiabatic path (the so-called annealing schedule) by the following time-dependent Hamiltonian

$$\boldsymbol{H}(t) = A(t)\boldsymbol{H}_i + B(t)\boldsymbol{H}_f \text{ with } t \in [0, t^{\max}], \tag{11}$$

[1] In quantum computing, $|\bullet\rangle$ is the Dirac notation to represent a unit vector in a multi-dimensional complex Hilbert space. More details can be found in [51, 52].

  where $t^{\max}$ is the annealing time and $A(t)$ and $B(t)$ are functions satisfying $A(0) = 1 \gg B(0)$ and $B(t^{\max}) = 1 \gg A(t^{\max})$. The details can be found in the review by [43].
- At the end of the annealing process, the ground state of the targeted Hamiltonian $\boldsymbol{H}_f$ is obtained upon measurement.

Ising Hamiltonian is widely used in the quantum annealing since Ising Hamiltonian-based quantum annealers are commercially available. Let us consider an undirected graph $(V, E)$ where $V = \{1, \ldots, K\}$ is a set of $K$ qubits and $E$ specifies the set of interactions between two qubits, *i.e.* $E \subset \{(i,j) | i \in V, j \in V, \text{and } i < j\}$, the Ising Hamiltonian [20–22] is written as

$$\boldsymbol{H}_f^{\text{Ising}} = \sum_{i \in V} h_i \boldsymbol{\sigma}_i^Z + \sum_{(i,j) \in E} J_{ij} \boldsymbol{\sigma}_i^Z \otimes \boldsymbol{\sigma}_j^Z \,, \tag{12}$$

where $\boldsymbol{\sigma}_i^Z$ is the Pauli-Z operator applied on qubit $i$, $\boldsymbol{\sigma}_i^Z \otimes \boldsymbol{\sigma}_j^Z$ are the ones applied on qubits $i$ and $j$, and $hi\ \forall i$ and $Jij\ \forall i,j$ are constants. As a result, $H_f^{\text{Ising}}$ is a $2^K \times 2^K$ diagonal matrix whose eigenvalues correspond to the following so-called Ising function

$$F_{\text{Ising}}(\boldsymbol{s}; \boldsymbol{h}, \boldsymbol{J}) = \sum_{i \in V} h_i s_i + \sum_{(i,j) \in E} J_{ij} s_i s_j \text{ with } s = [s_1 \ \ldots \ s_K]^T \in \{-1, 1\}^K. \tag{13}$$

In this equation, $\boldsymbol{s}$ corresponds to the spin vector and can be seen as the unknowns while $\boldsymbol{h} = [h_i \text{ for } i = 1, \ldots, K]$ and $\boldsymbol{J} = [J_{ij} \text{ for } i = 1, \ldots, K \text{ and } j = 1, \ldots, K]$ encode the Hamiltonian and are user-defined parameters. As a result, looking for the ground state of $\boldsymbol{H}_f^{\text{Ising}}$ corresponds to the minimisation of $F_{\text{Ising}}$ as

$$\boldsymbol{s} = \arg\min_{\boldsymbol{s}'} F_{\text{Ising}}(\boldsymbol{s}'; \boldsymbol{h}, \boldsymbol{J}) \,. \tag{14}$$

Eq. (14) can be efficiently solved with QA.

Using the transformation $b_i = \frac{1+s_i}{2} \in \{0,1\} \forall i$, with $b_i = b_i^2$, the Ising problem (14) can be transformed into a QUBO problem [44] as

$$\boldsymbol{b} = \arg\min_{\boldsymbol{b}'} F_{\text{QUBO}}(\boldsymbol{b}'; \boldsymbol{Q}) \,, \tag{15}$$

where $\boldsymbol{Q}$ is the QUBO matrix, which is constructed based on the user-defined problem parameters $\boldsymbol{h}$ and $\boldsymbol{J}$, and

$$F_{\text{QUBO}}(\boldsymbol{b};\boldsymbol{Q}) = \sum_{i=1}^{K}\sum_{j=1}^{K} Q_{ij} b_i b_j = \boldsymbol{b}^T \boldsymbol{Q} \boldsymbol{b}, \tag{16}$$

with $\boldsymbol{b} = [b_1 \ \dots \ b_K]^T \in \{0,1\}^K$ being the set of the binary problem unknowns.

Using an appropriate initial Hamiltonian $\boldsymbol{H}_i$, considering $\boldsymbol{H}_f = \boldsymbol{H}_f^{\text{Ising}}$, and a suitable annealing schedule (11), the quantum annealing algorithm allows finding the ground state of $\boldsymbol{H}_f^{\text{Ising}}$, and thus providing the solution of an Ising problem (14) or of a QUBO problem (15). The progress in developing quantum annealing hardware has been marked by several significant advancements over the past decade. In particular, the available D-Wave Systems [45] are applied in various applications [24].

## 4. Quantum annealing assisted structure optimisation

In this section, first we introduce a general sequential programming (SP) framework - a generalised version of the sequential quadratic programming [46] - in which an optimisation problem is transformed into a sequence of optimisation problems of polynomial forms by Taylor's series. Then the quantum annealing-assisted sequential programming (QA-SP) is detailed for quadratic polynomial forms. Finally, the developed QA-SP is employed to perform structure optimisation.

### *4.1 Sequential programming (SP)*

Sequential programming (SP), in general, is an iterative strategy to solve a complex optimisation problem, in which a sequence of less complex optimisation problems is performed to reach the optimum solution. Let us consider the following quasi-unconstrained minimisation problem

$$\begin{cases} \min\limits_{\boldsymbol{x}} f(\boldsymbol{x}) \\ \text{subject to } \boldsymbol{x}_{\min} \le \boldsymbol{x} \le \boldsymbol{x}_{\max} \end{cases}, \tag{17}$$

where $f(\boldsymbol{x})$ is a multivariate function $f : \mathbb{R}^N \to \mathbb{R}$ of the argument $\boldsymbol{x}$ and $\boldsymbol{x}_{\min}$ and $\boldsymbol{x}_{\max}$ are respectively its lower and upper bounds. We assume that, in the neighbourhood of $\boldsymbol{x}$, there exists a Taylor's series expansion up to the $P^{th}$ order with $P \ge 1$ as[2]

[2] For any matrix/vector $\boldsymbol{c}$, one defines a Kronecker product power $\boldsymbol{c}^{\otimes m} = \underbrace{\boldsymbol{c} \otimes \dots \otimes \boldsymbol{c}}_{\boldsymbol{c} \text{ is repeated } m \text{ times}}$

$$f(\boldsymbol{x}+\boldsymbol{\delta}) = f(\boldsymbol{x}) + \sum_{m=1}^{P} \frac{1}{m!} \left( \frac{\partial^m f(\boldsymbol{x})}{\partial \boldsymbol{x}^{\otimes m}} \right)^T \boldsymbol{\delta}^{\otimes m} + \mathcal{O}(|\boldsymbol{\delta}|^{P+1}). \quad (18)$$

Consequently, one can define the polynomial approximation of a function $f$ in the neighbourhood of $\boldsymbol{x}$ by

$$f(\boldsymbol{x}+\boldsymbol{\delta}) \approx \Phi^P(\boldsymbol{\delta}; \boldsymbol{x}, f) = f(\boldsymbol{x}) + \sum_{m=1}^{P} \frac{1}{m!} \left( \frac{\partial^m f(\boldsymbol{x})}{\partial \boldsymbol{x}^{\otimes m}} \right)^T \boldsymbol{\delta}^{\otimes m}, \quad (19)$$

where $\Phi^{\bullet}(\bullet;\bullet)$ denotes the polynomial approximation operator.

The key step in the SP strategy is to iteratively construct the approximated form of the problem (17) using Taylor's series (19) at the current guess, *i.e.* $\boldsymbol{x}^{(k)}$ with $k \geq 0$, to find the next guess, *i.e.* $\boldsymbol{x}^{(k+1)}$. Assuming $\boldsymbol{x}^{k+1} = \boldsymbol{x}^k + \boldsymbol{\delta}^{k+1}$, $\boldsymbol{\delta}^{k+1}$ is then obtained by solving the following minimisation problem

$$\begin{cases} \boldsymbol{\delta}^{(k+1)} = \arg\min_{\boldsymbol{\delta}} \Phi^P\left(\boldsymbol{\delta}; \boldsymbol{x}^{(k)}, f\right) \\ \text{subject to} \max(\boldsymbol{x}_{\min} - \boldsymbol{x}^{(k)}, \boldsymbol{\delta}^b_{\min}) \leq \boldsymbol{\delta} \leq \min(\boldsymbol{x}_{\max} - \boldsymbol{x}^{(k)}, \boldsymbol{\delta}^b_{\max}) \end{cases}, \quad (20)$$

where $\boldsymbol{\delta}^b_{\min} \leq \boldsymbol{0}$ and $\boldsymbol{\delta}^b_{\max} \geq 0$ define a hyper-box around $\boldsymbol{x}^{(k)}$ due to the fact that the polynomial approximation (19) is generally accurate only within a small region around $\boldsymbol{x}$. Besides, the lower bound $\boldsymbol{x}_{\min}$ and the upper bound $\boldsymbol{x}_{\max}$ are taken into account, resulting in the largest box defined by $\boldsymbol{x}_{\min} - \boldsymbol{x}^{(k)} \leq 0$ and $\boldsymbol{x}_{\max} - \boldsymbol{x}^{(k)} \geq 0$. As a result, the intersection of these two boxes leads to the bound constraints in Eq. (20). The iterative procedure of the SP strategy is summarised in Algorithm 1. In this work, the polynomial optimisation problem (20) is solved by a hybrid algorithm combining quantum computing and classical computers, as detailed in the next section.

Algorithm 1 - Iterative procedure of the SP strategy for the minimisation problem (17) incorporating the polynomial approximation (19) at each step.

1) initialise to an admissible initial solution $\boldsymbol{x}^{(0)} \in [\boldsymbol{x}_{\min}, \boldsymbol{x}_{\max}]$;
2) set the box bounds $\boldsymbol{\delta}^b_{\min}$ and $\boldsymbol{\delta}^b_{\max}$;
3) set $k = 0$;
4) **repeat**
5) solve the optimisation problem (20) to obtain $\boldsymbol{\delta}^{(k+1)}$;
6) update $\boldsymbol{x}^{k+1} = \boldsymbol{x}^k + \boldsymbol{\delta}^{k+1}$;
7) $k \leftarrow k + 1$;
8) **until** convergence;

When the minimisation problem (17) is incorporated with equality and inequality constraints as stated as follows:

$$\min_{\boldsymbol{x}} f(\boldsymbol{x}) \;\text{ subject to} \begin{cases} h_j(\boldsymbol{x}) = 0 \text{ with } j = 1, \dots, N_h, \\ l_j(\boldsymbol{x}) \leq 0 \text{ with } j = 1, \dots, N_l, \text{and} \\ \boldsymbol{x}_{\min} \leq \boldsymbol{x} \leq \boldsymbol{x}_{\max} \end{cases}, \tag{21}$$

where $N_h$ and $N_l$ are the number of equality and inequality constraints, respectively, the corresponding constrained minimisation problem is transformed into the canonical form (20) using penalties. For this purpose, an augmented polynomial minimisation form is defined by

$$f_{\text{aug}}(\boldsymbol{x}, \boldsymbol{\lambda}) = f(\boldsymbol{x}) + \sum_{j=1}^{N_h} c_j^h \left[h_j(\boldsymbol{x})\right]^2 + \sum_{j=1}^{N_l} c_j^l \left[l_j(\boldsymbol{x}) + \lambda_j\right]^2, \tag{22}$$

where $c_j^h$ and $c_j^l$ are the penalty factors associated respectively with the equality constraint $h_j$ and with inequality constraint $l_j$, and $\boldsymbol{\lambda} = [\lambda_1 \dots \lambda_{N_l}]$ is a vector of the auxiliary variables which are constrained to be non-negative to enforce the inequality constraints. The values $c_j^h$ and $c_j^l$ $\forall j$ need to be large enough so that the optimization solution does not depend on this choice. Indeed, the solution of the minimisation problem satisfies exactly the constraints. As a result, the constrained form (21) can be rewritten in the quasi-unconstrained form (17) as

$$\begin{cases} \min\limits_{x,\lambda} f(\boldsymbol{x}, \boldsymbol{\lambda}) \\ \text{subject to } \boldsymbol{x}_{\min} \leq \boldsymbol{x} \leq \boldsymbol{x}_{\max} \text{ and } 0 \leq \boldsymbol{\lambda} \leq +\infty \end{cases}, \tag{23}$$

Finally, the minimisation problem (21) can be solved with Algorithm *1* using the augmented polynomial approximation (23) in the approximated polynomial form (20).

In practice, careful calibration of the penalty coefficients in Eq. (22) is crucial to obtain an appropriate balance between the objective functions and constraints. Excessively large coefficients may lead to an over-contribution of the constraints, potentially overshadowing the objective function and restricting the exploration of feasible solutions. Conversely, too small coefficients could result in the constraints being ignored, leading to solutions that do not satisfy the required conditions. As shown in Section 4.2, the augmented form (22) is used to construct the QUBO problem, so the choice of penalty values directly influences the performance of the quantum annealing algorithms. To ensure effective performance and avoid undesirable outcomes, penalty values must be chosen within an appropriate range - neither too large nor too small [44, 47].

### *4.2 Quantum annealing-assisted sequential programming (QA-SP)*

When the polynomial forms, e.g. Eq. (19), are quadratic, QUBOs are directly obtained by approximating each continuous variable with binary variables through a linear transformation. The strategies for the continuous-binary and quadratic polynomial form -QUBO approximation, along with a nested hybrid QA-SP scheme for Algorithm 1, are detailed in the following.

#### 4.2.1 Approximating continuous variables with binary variables

In general, a real value $\delta$ can be approximated by binary variables under the following general form

$$\delta(b_1,\dots,b_L) \equiv \sigma + \kappa \sum_{j=1}^{L} \zeta_j b_j\,, \tag{24}$$

where $\sigma$ and $\kappa$ are the respectively offset and scaling values respectively, and $\xi_j$ with $j = 1,\dots,L$ are constants. There are different ways to construct such a continuous-to-binary approximation, depending on how the constants $\zeta_j$ with $j = 1,\dots,L$ are chosen. For example, in the random representation, $\zeta_j, \forall j$, are randomly generated [34, 40], in the uniform representation, $\zeta_j, \forall j$, take the same value, as considered in [36], and the binary representation considers $\zeta_j = 2^{j-1}, \forall j$, see e.g. in [30, 31, 39]. These strategies were investigated in [40], showing that the accuracy of the solution varies significantly depending on how the real numbers are represented and the advantages and disadvantages of these representations depend on the problem size and the difficulty of the problem being solved. In this work, we adopt the binary approximation, as it provides the highest representational capability [40]. As a result, Eq. (24) can be rewritten as

$$\delta\colon (b_1,\dots b_N) \rightarrow\ \bar{\delta} + \epsilon\left(\sum_{j=0}^{L-1} b_{j+1}\, 2^j - 2^{L-1} + 1\right) \in [d_{\min}, d_{\max}]. \tag{25}$$

where $\epsilon$ is the so-called discretisation error, $\bar{\delta}$ is the central value, and where

$$d_{\min} = \bar{\delta} - (2^{L-1} - 1)\epsilon \text{ and } d_{\max} = \bar{\delta} + 2^{L-1}\epsilon. \tag{26}$$

Eq. (25) implies that a continuous variable δ is discretised into $2^L$ discrete values in the range of $[d_{\min}, d_{\max}]$. The last equation allows estimating the pair $(\bar{\delta}, \epsilon)$ from the bound $[d_{\min}, d_{\max}]$ as

$$\epsilon = \frac{d_{\max} - d_{\min}}{2^L - 1} \text{ and } \bar{\delta} = \frac{d_{\max} + d_{\min} - \epsilon}{2}. \tag{27}$$

A larger range $[d_{\min}, d_{\max}]$ is obtained using more qubits for a given error $\epsilon$, whilst when the number of qubits is fixed, the range $[d_{\min}, d_{\max}]$ becomes smaller with decreasing $\epsilon$. Eq. (25) can be rewritten as

$$\delta(\boldsymbol{b}) = d_{\min} + \epsilon \boldsymbol{\beta}^T \boldsymbol{b}, \tag{28}$$

where $\boldsymbol{\beta} = [2^0\ 2^1 \dots 2^{L-1}]^T$ and $\boldsymbol{b} = [b_1 \dots b_L]^T$.

Let us consider a vector $\boldsymbol{\delta}$ of $W$ components, Eq. (28) is thus applied to each component of $\boldsymbol{\delta}$, leading to

$$\boldsymbol{\delta} \approx \boldsymbol{d}_{\min} + [\epsilon_i \boldsymbol{\beta}^T \boldsymbol{b}_i \text{ for } i = 1, \dots, W] = \boldsymbol{d}_{\min} + \boldsymbol{V}\boldsymbol{b} \in [\boldsymbol{d}_{\min}, \boldsymbol{d}_{\max}], \tag{29}$$

where $\bullet_i$ denotes a quantity associated with the component $\delta_i \in \boldsymbol{\delta}$, $\boldsymbol{d}_{\min}$ and $\boldsymbol{d}_{\max}$ denote respectively the lower-bound and upper-bound of $\boldsymbol{\delta}$, and where

$$\boldsymbol{d}_{\min} = \left[d_{\min,1} \dots d_{\min,W}\right]^T, \tag{30}$$

$$\boldsymbol{V} = \operatorname{diag}(\epsilon_1 \boldsymbol{\beta}^T, \dots, \epsilon_W \boldsymbol{\beta}^T), \text{and} \tag{31}$$

$$\boldsymbol{b} = [\boldsymbol{b}_1^T \dots \boldsymbol{b}_W^T]^T, \tag{32}$$

with $\boldsymbol{b}$ being the vector of all binary variables and diag($\bullet$) being a diagonal matrix of $W$ blocks. We can also denote

$$\boldsymbol{\epsilon} = [\epsilon_1 \dots \epsilon_W]^T, \tag{33}$$

as the discretisation error vector.

### 4.2.2 Approximating a quadratic polynomial form as a QUBO

Using the linear continuous-binary conversion (29) of $\boldsymbol{\delta}$ in the polynomial forms, *e.g*. Eq. (19), leads to a binary polynomial form of the same degree. We consider Eq. (19) under a quadratic form which is achieved by considering linear or quadratic Taylor's series for the objective function and for constraints when present. These quadratic forms are rewritten in the general form

$$\boldsymbol{P}(\boldsymbol{\delta}) = \boldsymbol{\delta}^T \boldsymbol{g} + \boldsymbol{\delta}^T \boldsymbol{A} \boldsymbol{\delta}, \tag{34}$$

where $g = \frac{\partial f}{\partial \boldsymbol{x}}$ is the linear coefficient vector, $A = \frac{1}{2}\frac{\partial^2}{\partial \boldsymbol{x} \partial \boldsymbol{x}}$ is the quadratic coefficient matrix, which is symmetric, and where the constant term is omitted as it is irrelevant in this minimisation context. By replacing $\boldsymbol{\delta}$ in Eq. (34) by Eq. (29), one has

$$\Lambda(\boldsymbol{b}) = \boldsymbol{d}_{\min}^T \boldsymbol{g} + \boldsymbol{d}_{\min}^T \boldsymbol{A} \boldsymbol{d}_{\min} + \boldsymbol{b}^T \boldsymbol{V}^T (\boldsymbol{g} + \boldsymbol{2} \boldsymbol{A} \boldsymbol{d}_{\min}) + \boldsymbol{b}^T \boldsymbol{V}^T \boldsymbol{A} \boldsymbol{V} \boldsymbol{b}. \tag{35}$$

Since $b_i^2 = b_i \ \forall i$ for $b_i \in \{0,1\}$, Eq. (35) can be rewritten as

$$\Lambda(\boldsymbol{b}) = \boldsymbol{d}_{\min}^T \boldsymbol{g} + \boldsymbol{d}_{\min}^T \boldsymbol{A} \boldsymbol{d}_{\min} + F_{\text{QUBO}}(\boldsymbol{b}, \boldsymbol{Q}), \tag{36}$$

where $F_{\text{QUBO}}$is the QUBO form (15) with the QUBO matrix

$$\boldsymbol{Q} = \boldsymbol{V}^T \boldsymbol{A} \boldsymbol{V} + \operatorname{diag}\left(\boldsymbol{V}^T (\boldsymbol{g} + 2 \boldsymbol{A} \boldsymbol{d}_{\min})\right), \tag{37}$$

where diag($\boldsymbol{a}$) is the diagonal matrix, whose diagonal is the vector $\boldsymbol{a}$. As a result, the minimisation of the polynomial form (34) under constraints can be carried out through solving the QUBO (36) with quantum annealers as referred to Eq. (15).

#### 4.2.3 A nested hybrid framework

The hybrid quantum annealing-assisted sequential programming (QA-SP) algorithm is summarised in Algorithm 2, in which the part working on quantum annealers is highlighted while the remaining part is carried out with classical computers. The algorithm does not require solving any linear systems. Although quantum annealers are susceptible to various types of noise and errors, the iterative procedure is designed to tolerate these imperfections. Since each iteration only requires an approximate solution to the QUBO, the method continues to converge toward the optimal solution within a finite number of iterations.

At the iteration $k$ with $\boldsymbol{x}^{(k)}$ being known, the quadratic polynomial approximations (19) of the minimisation problem is constructed. The side constraints of Eq. (20) are also re-evaluated with respect to the box $\left[\boldsymbol{\delta}_{\min}^{b}, \boldsymbol{\delta}_{\max}^{b}\right]$ and the bounds $[\boldsymbol{x}_{\min}, \boldsymbol{x}_{\max}]$ of the variables. A QUBO problem is obtained subsequently using the continuous-binary approximation (29) and can be solved directly with quantum annealers. After updating the increment $\boldsymbol{\delta}^{(k+1)}$ with the solution $\boldsymbol{b}^{(k+1)}$ from quantum annealers following (29), two scenarios are envisioned:

- The objective function value decreases as expected the unknown $\boldsymbol{x}^{(k+1)}$ is encoded and the polynomial approximation at this point is constructed for the next iteration.
- The objective function value does not decrease. The continuous-binary approximation needs to be refined by reducing the discretisation error $\boldsymbol{\epsilon}$ (see Eq. (33)) by a so-called shrinking factor $\xi < 1$. We consider $\xi = 0.5$ in this work.
- After $N_{\text{success}}$ consecutive iterations in which the objective function increases, the discretisation error $\boldsymbol{\epsilon}$ increases by a factor of $\xi^{-1}$. We consider $N_{\text{success}} = 50$ in this work.

We note that even for quadratic functional, an iterative procedure is required. On the one hand, it is needed because of the binarisation process with a low number of qubits. On the other hand, it allows accounting for the error inherent to the quantum annealing process.

In Algorithm 2, the box defined by $\boldsymbol{\delta}_{\min}^{b}$ and $\boldsymbol{\delta}_{\max}^{b}$ serves as a safeguard to maintain the quality of the Taylor's approximation. In addition, the discretisation error $\boldsymbol{\epsilon}$ is introduced as a user parameter to control the accuracy of approximating continuous variables by binary values, as detailed in Section 4.2.1. For a given pair $(\boldsymbol{\epsilon}, \overline{\boldsymbol{\delta}})$, the actual bounds $\boldsymbol{d}_{\min}$ and $\boldsymbol{d}_{\min}$ are estimated (line 13 in Algorithm 2), and they are then limited by the box size (line 14 in Algorithm 2). Both the box and the discretisation error are problem dependent. In the course of the optimisation iterations, the discretisation error is gradually reduced (line 26 in Algorithm 2) to refine the solution's location. In our applications in Section 5, a box $[\boldsymbol{\delta}_{\min}^{b}, \boldsymbol{\delta}_{\max}^{b}]$ of $[-0.05, 0.05]$ and an initial discretisation error of $\frac{10^{-4}}{2^{L}-1}$, where $L$ represents the number of qubits per continuous unknown, are chosen for each continuous unknown.

On current quantum hardware, the available qubits and their connectivity are limited. Additionally, errors in the quantum machine become more pronounced as the number of qubits increases. Therefore, at this stage, it is preferable to use as few qubits as possible. Since the SP strategy described in Algorithm 2 considers an incremental update, *i.e.* $\boldsymbol{x}^{(k+1)} \leftarrow \boldsymbol{x}^{(k)} + \boldsymbol{\delta}^{(k+1)}$, each component of $\boldsymbol{\delta}^{(k+1)}$ requires at least three possibilities: negative, zero, and positive to move towards the optimised solution. Therefore, the binary approximation (25) needs at least two qubits ($L = 2$) per variable, in which case Eq. (25) becomes

$$\delta(b_1, b_2) = \bar{\delta} + \epsilon(b_1 + 2b_2 - 1) \in \{\bar{\delta} - \epsilon, \bar{\delta}, \bar{\delta} + \epsilon, \bar{\delta} + 2\epsilon\}, \tag{38}$$

The incremental update above favours increasing the variable over decreasing it. In the remaining parts of the paper, we consider two qubits for each continuous variable.

Algorithm 2 - Quantum annealing-assisted sequential programming (QA-SP) algorithm.

1) set number of qubits per continuous variable $L$;
2) set initial discretisation error $\boldsymbol{\epsilon}_0$;
3) set box $\boldsymbol{\delta}^b_{\min}$, $\boldsymbol{\delta}^b_{\max}$;
4) set shrinking factor $\xi$;
5) set value of $N_{\text{steps}}$;
6) set values of $N_{\text{failed}}$ and of $N_{\text{success}}$;
7) initialise admissible initial solution $\boldsymbol{x}^{(0)} \in [\boldsymbol{x}_{\min}, \boldsymbol{x}_{\max}]$ ;
8) construct polynomial approximation around $\boldsymbol{x}^{(0)}$ as discussed in Section 4.1;
9) $k \leftarrow 0$, $r \leftarrow 0$, and $c \leftarrow 0$;
10) $\boldsymbol{\epsilon} \leftarrow \boldsymbol{\epsilon}_0$ and $\bar{\boldsymbol{\delta}} \leftarrow \mathbf{0}$;
11) **repeat**
12) update bounds with respect to the box

$$\boldsymbol{\delta}^{(k)}_{\min} \leftarrow \max(\boldsymbol{x}_{\min} - \boldsymbol{x}^{(k)}, \delta^b_{\min}) \text{ and } \boldsymbol{\delta}^{(k)}_{\max} \leftarrow \min\left(\boldsymbol{x}_{\max} - \boldsymbol{x}^{(k)}, \delta^b_{\max}\right); \tag{39}$$

13) estimate discretisation range $[\boldsymbol{d}_{\min}, \boldsymbol{d}_{\max}]$ following Eq. (26);
14) update discretisation range with respect to bounds

$$\boldsymbol{d}^{(k)}_{\min} \leftarrow \max(\boldsymbol{d}_{\min}, \boldsymbol{\delta}^{(k)}_{\min}) \text{ and } \boldsymbol{d}^{(k)}_{\max} \leftarrow \min(\boldsymbol{d}_{\max}, \boldsymbol{\delta}^{(k)}_{\max}); \tag{40}$$

15) update $\boldsymbol{\epsilon}$ and $\bar{\boldsymbol{\delta}}$ with new $\boldsymbol{d}_{\min}$ and $\boldsymbol{d}_{\max}$ following Eq. (27);
16) evaluate the continuous-binary approximation (29) as discussed in Section 4.2.1;
17) estimate the QUBO matrix following Section 4.2.2;
18) call quantum annealers (15) leading to the solution $\boldsymbol{b}^{(k+1)}$;
19) update $\boldsymbol{\delta}^{(k+1)}$ with $\boldsymbol{b}^{(k+1)}$ following Eq. (29);
20) **if** $f\left(\boldsymbol{x}^{(k)} + \boldsymbol{\delta}^{(k+1)}\right) < f\left(\boldsymbol{x}^{(k)}\right)$ **then**
21) $\boldsymbol{x}^{(k+1)} \leftarrow \boldsymbol{x}^{(k)} + \boldsymbol{\delta}^{(k+1)}$;
22) $c \leftarrow c + 1$; and $k \leftarrow k + 1$;
23) construct polynomial approximation around $\boldsymbol{x}^{(k)}$ as discussed in Section 4.1;
24) **else**
25) $c \leftarrow 0$; and $r \leftarrow r + 1$;
26) $\boldsymbol{\epsilon} \leftarrow \max(\boldsymbol{\epsilon}_{\min}, \xi\boldsymbol{\epsilon})$;
27) **end if**
28) **if** $k \geq N_{\text{steps}}$ or $r \geq N_{\text{failed}}$ **then**
29) break;
30) **end if**
31) **if** $c \geq N_{\text{success}}$ **then**
32) $c \leftarrow 0$; and $\boldsymbol{\epsilon} \leftarrow \min(\boldsymbol{\epsilon}_0, \boldsymbol{\epsilon}/\xi)$;
33) **end if**
34) **until** convergence;

### *4.3 Application for optimisation of truss structures*

In this section, the hybrid framework described in Algorithm 2 is considered for the optimisation of truss structures. The optimisation process involves multiple iterations. At each iteration, the state variables U and the design variables α are updated through two distinct minimisations. At the $k+1^{th}$ iteration, their values at the previous step, *i.e.*, $\boldsymbol{U}^{(k)}$ and $\boldsymbol{\alpha}^{(k)}$, are known, and the values at the current step, *i.e.*, $\boldsymbol{U}^{(k+1)}$ and $\boldsymbol{\alpha}^{(k+1)}$, are sought. These two minimisations are sequentially carried out as follows:

- The value of $\boldsymbol{U}^{(k+1)}$ is first obtained by solving the unconstrained minimisation (7) which is rewritten as
$$\boldsymbol{U}^{(k+1)}\left(\boldsymbol{\alpha}^{(k)}\right) = \arg\min_{\boldsymbol{U}} \Psi\left(\boldsymbol{U};\boldsymbol{\alpha}^{(k)}\right). \tag{41}$$
This minimisation problem will be solved by Algorithm 2 using a quadratic Taylor's series of the objective function.
- The value of $\boldsymbol{\alpha}^{(k+1)}$ is obtained by solving the constrained minimisation (9), in which one key element is to estimate the global compliance $C(\boldsymbol{\alpha})$. In this step, instead of finding $C(\boldsymbol{\alpha})$ from the solution of the minimisation problem (7), a linear Taylor's series is employed since the estimation of the hessian in the quadratic form is challenging in this case, leading to
$$C(\boldsymbol{\alpha}) \approx \Phi^1\left(\boldsymbol{\alpha}-\boldsymbol{\alpha}^{(k)};\boldsymbol{\alpha}^{(k)},C\right) = \boldsymbol{F}^T\boldsymbol{U}^{(k+1)} + \boldsymbol{\omega}^{(k)T}\left(\boldsymbol{\alpha}-\boldsymbol{\alpha}^{(k)}\right), \tag{42}$$
where both $\boldsymbol{U}^{(k+1)}$ and $\boldsymbol{\omega}^{(k)} = \frac{\partial\left(\boldsymbol{F}^T\boldsymbol{U}^{(k+1)}\right)\left(\boldsymbol{\alpha}^{(k)}\right)}{\partial\boldsymbol{\alpha}^{(k)}}$ are known from the solutions of the problem (41). Moreover, the constraints in the problem are also relaxed by applying a first-order Taylor's series. As a result, Eq. (9) is rewritten as
$$\boldsymbol{\alpha}^{(k+1)} = \boldsymbol{\alpha}^{(k)} + \arg\min_{\boldsymbol{\delta}} \Phi^1\left(\boldsymbol{\delta},\boldsymbol{\alpha}^{(k)},C\right), \tag{43}$$
$$\text{subject to } \begin{cases} \Phi^1\left(\boldsymbol{\delta},\boldsymbol{\alpha}^{(k)},h\right) = h\left(\boldsymbol{\alpha}^{(k)}\right) + \boldsymbol{D}^{(k)T}\boldsymbol{\delta} = 0, \\ \boldsymbol{\alpha}_{\min} - \boldsymbol{\alpha}^{(k)} \leq \boldsymbol{\delta} \leq \boldsymbol{\alpha}_{\max} - \boldsymbol{\alpha}^{(k)}, \end{cases}$$
where $\Phi^1\left(\cdot,\boldsymbol{\alpha}^{(k)},f\right)$ denotes a linear approximation operator of a function $f$ in the neighbourhood of $\boldsymbol{\alpha}^{(k)}$, $h(\boldsymbol{\alpha}) = \sum_{k=1}^{N}\alpha_k L_k A_k^0 - V^{\text{target}}$, $\boldsymbol{\delta} = \boldsymbol{\alpha}^{(k+1)} - \boldsymbol{\alpha}^{(k)}$ is the increments of design variables, and $\boldsymbol{D}^{(k)} = \frac{\partial h}{\partial\boldsymbol{\alpha}^{(k)}}$ is the constraint gradient. This minimisation problem is also solved by Algorithm *2*.

To estimate the sensitivity of $\boldsymbol{U}^{(k+1)}$ with respect to $\boldsymbol{\alpha}^{(k)}$, *i.e.* to evaluate $\boldsymbol{\omega}^{(k)} = \frac{\partial\left(\boldsymbol{F}^T\boldsymbol{U}^{(k+1)}\right)}{\partial\boldsymbol{\alpha}^{(k)}}$ after solving Eq. (41), we consider the consistency condition

$$\frac{\partial}{\partial\alpha_i^{(k)}}\left(\boldsymbol{K}\left(\boldsymbol{\alpha}^{(k)}\right)\boldsymbol{U}^{(k+1)}\right) = \boldsymbol{0}, \tag{44}$$

leading to

$$\frac{\partial \boldsymbol{U}^{(k+1)}}{\partial \alpha_i^{(k)}} = -\boldsymbol{K}\left(\boldsymbol{\alpha}^{(k)}\right)^{-1} \frac{\partial \boldsymbol{K}\left(\boldsymbol{\alpha}^{(k)}\right)}{\partial \alpha_i^{(k)}} \boldsymbol{U}^{(k+1)}, \text{and} \tag{45}$$

$$\boldsymbol{F}^T \frac{\partial \boldsymbol{U}^{(k+1)}}{\partial \alpha_i^{(k)}} = -\boldsymbol{U}^{(k+1)T} \frac{\partial \boldsymbol{K}\left(\boldsymbol{\alpha}^{(k)}\right)}{\partial \alpha_i^{(k)}} \boldsymbol{U}^{(k+1)}. \tag{46}$$

The last equation allows estimating each component $\omega_i^{(k)}$ of $\boldsymbol{\omega}^{(k)}$ as

$$\omega_i^{(k)} = -\boldsymbol{U}^{(k+1)T} \frac{\partial \boldsymbol{K}\left(\boldsymbol{\alpha}^{(k)}\right)}{\partial \alpha_i^{(k)}} \boldsymbol{U}^{(k+1)} = -\boldsymbol{U}^{(k+1)T} \boldsymbol{K}_i \boldsymbol{U}^{(k+1)}, \tag{47}$$

where $\boldsymbol{K}_i$ is obtained from $\boldsymbol{K}$ following Eq. (5).

The resolution scheme of the minimisation problem (9) is sketched in Algorithm 3, in which the QA-SP detailed in Algorithm *2* is used to solve each minimisation problem at each iteration. Starting from the initial structure characterised by $\boldsymbol{\alpha}^{(0)}$, the QA-SP algorithm in Algorithm 2 is used to solve the mechanical boundary value problem and its solution is then used to approximate the second minimization for the structure design update. These two minimisation problems are repeated until convergence.

Algorithm 3 - Iterative scheme for optimisation of truss structure.

1) initialise to an admissible initial solution $\boldsymbol{\alpha}^{(0)} \in [\boldsymbol{\alpha}_{\min}, \boldsymbol{\alpha}_{\max}]$;
2) set $k = 0$;
3) **repeat**
4) solve the optimisation problem (41) with Algorithm 2 and return $\boldsymbol{U}^{(k+1)}$ and $\boldsymbol{\omega}^{(k)}$;
5) solve the optimisation problem (43) with Algorithm 2 and return $\boldsymbol{\alpha}^{(k+1)}$;
6) $k \leftarrow k + 1$;
7) **until** convergence;

As shown in Algorithm 3, each optimisation iteration requires first solving the structural balance equations (line 4) and then updating the design variables based on the design functionals estimated from the former (line 5). These steps are both considered as minimization problems, resulting in two interdependent minimization tasks. To accelerate the resolution of each minimization, parallel computing can be employed in the parts considering the classical computers to estimate different elementary matrices required to construct the QUBO matrix before sending it to the quantum annealers.

Algorithm 3 follows the general strategy in structural topology optimisation, where the mechanical balance equations are solved first, followed by an update of the design variables based on the sensitivity of the objective function and constraints with respect to these variables. Although this decoupling strategy may lead to a locally optimal solution since treating the two steps separately can result in a design that is optimal at each stage but not globally optimal, it offers practical advantages. From an engineering perspective, this approach typically yields a design that is significantly improved over the initial configuration.

## 5. Numerical examples

This section provides some benchmarks in both two-dimensional and three-dimensional cases to highlight the capability of our proposed framework for truss structure optimisation. The QUBO problem (15) is solved with the D-Wave system [48] using the python script reported in Algorithm 4. *DWaveSampler* allows submitting the problem to an available quantum processing unit while *EmbeddingComposite* is used to embed the QUBO into the hardware graph since the problem does not automatically fit in. The sampling process is carried out with 200 reads and an annealing time of 20 μs. The output is chosen as the variable that makes the QUBO the smallest. In the remaining of this section, we consider 2 qubits (*i.e.* $L = 2$) for each continuous variable when employing Algorithm *2*.

Additionally, the results are compared with those obtained using SA-SP, for which simulated annealing (SA) is available in the D-Wave software development kit [48]. In that case, the Python script reported in Algorithm 5 is employed instead of quantum annealing (QA) to solve the QUBO problem. Both QA-based and SA-based frameworks follow the algorithm presented in Algorithm 2, with the only difference being the solver used to solve the QUBO problem in line 18: SA for the classical approach and QA for the hybrid approach. The comparison also includes results obtained with the method of moving asymptotes (MMA) [49] available with the Python package [50], which is a widely used optimiser in structural optimisation.

Algorithm 4 - QUBO problem submitted to the D-Wave system.

```
from dwave.system import DWaveSampler, EmbeddingComposite

def solve_QUBO(Q):
  sampler=EmbeddingComposite(DWaveSampler())
  sampleset=sampler.sample_qubo(Q,num_reads=200,annealing_time=20)
  b=sampleset.first.sample
  return b
```

Algorithm 5 - Solving QUBO problem with simulated annealing (SA).

```
from dwave.samplers import SimulatedAnnealingSampler

def solve_QUBO_SA(Q):
  sampler= SimulatedAnnealingSampler()
  sampleset=sampler.sample_qubo(Q,num_reads=200)
  b=sampleset.first.sample
  return b
```

### *5.1 Two-dimensional cases*

We consider the three cases reported in [36], whose initial configurations and boundary conditions are reported in Figure 1. The Young's modulus is equal to 200 GPa. The initial cross-sectional area is set to 0.5 $m^2$ for all bars.

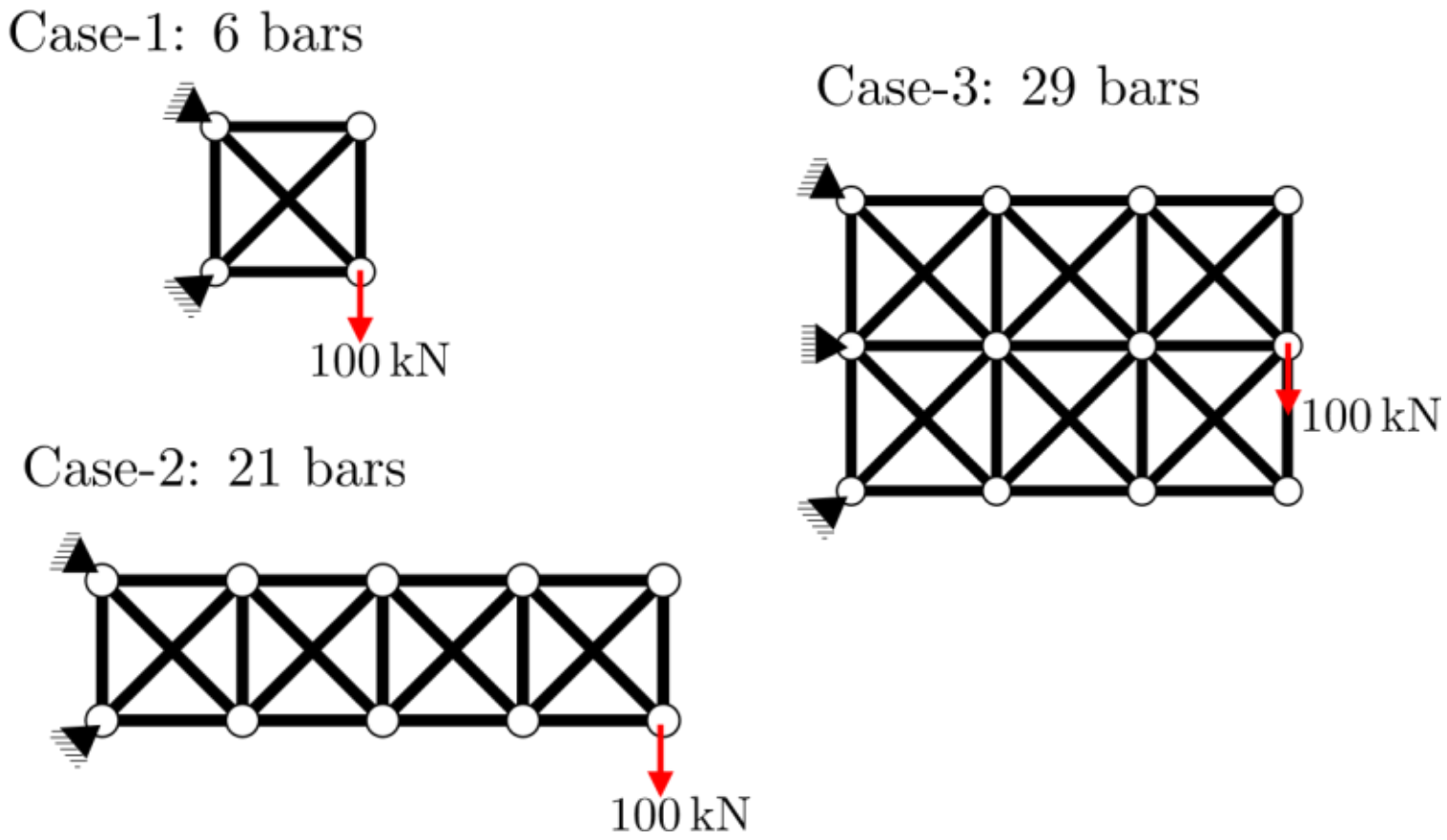

Figure 1 - Two-dimensional cases: three case studies of truss optimisation respectively with 6, 21, and 29 bars. The length is set to 1 m for both horizontal and vertical bars and to $\sqrt{2}$ m for the diagonal bars. The initial cross-sectional area is set to 0.5 $m^2$ for all these bars.

The capability of the proposed framework to solve a mechanical boundary value problem is first demonstrated through a minimisation of the potential energy as described by Eq. (7). The initial solution $\boldsymbol{U} = \boldsymbol{0}$, the initial discretisation error $\boldsymbol{\epsilon}_0 = \frac{10^{-4}}{2^L-1}[1 \dots 1]$ with $L$ being the number of qubits per continuous unknown and $\overline{\boldsymbol{\delta}} = \boldsymbol{0}$ are considered. The values $\boldsymbol{\epsilon}_0$ and $\overline{\boldsymbol{\delta}}$ allows estimating the initial discretisation range $[\boldsymbol{d}_{\min}, \boldsymbol{d}_{\max}]$ as discussed in Section 4.2.1. Each case is solved three times to assess the randomness in the resolution because of the heuristic nature of quantum annealers. The convergence histories of the three cases shown in Figure 1 are reported in Figure 2 where the exact value $\Psi^{\text{exact}}$ is estimated from $\boldsymbol{U}$ obtained by directly solving Eq. (3). The resolution of the case with 6 bars exhibits no discrepancy because of the relatively small number of qubits used in the optimisation process. For all the studied cases, the accuracy keeps improving with the number of iterations and a relative error as low as $10^{-9}$ can be reached in all cases.

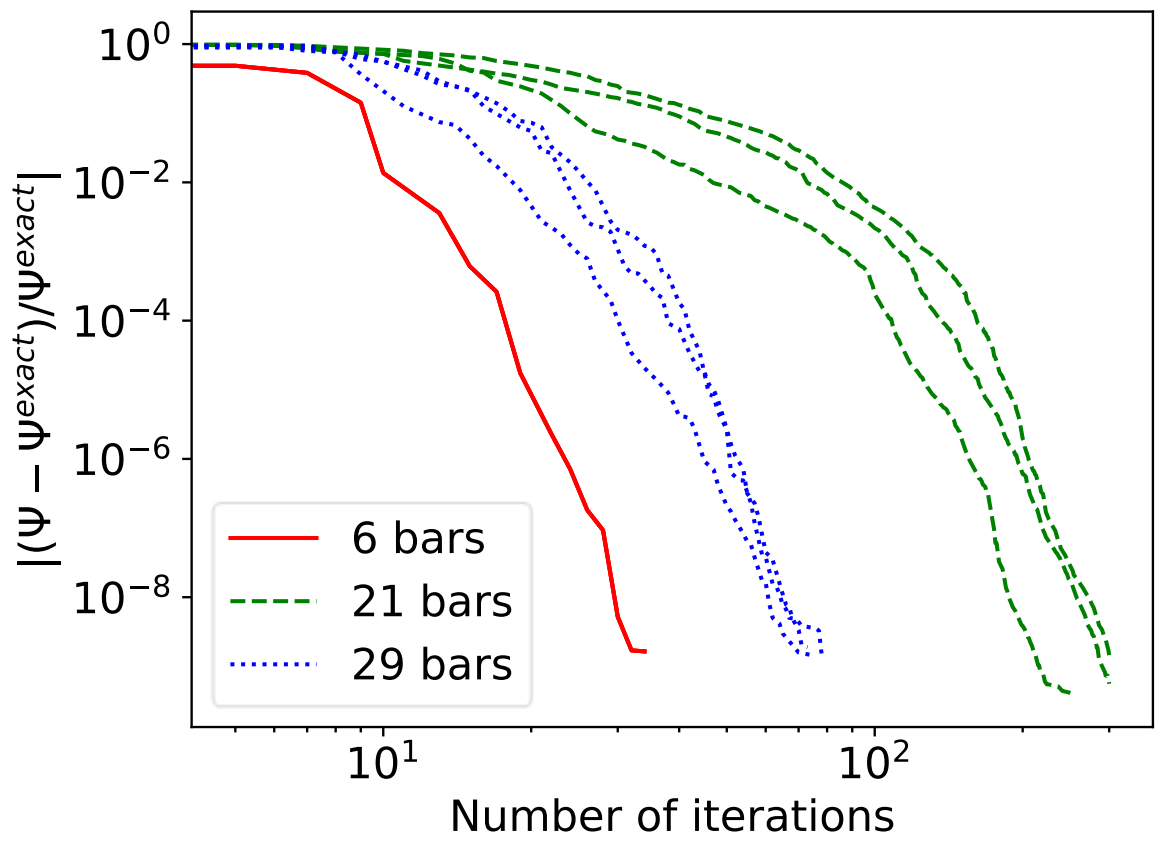

Figure 2 - Solving the equilibrium balance equation (7) with quantum annealers. The exact solution $\Psi^{exact}$ is obtained by directly solving Eq. (3).

First, we consider applying Algorithm 3 for the case of 6 bars. The QA-SP parameters reported above are chosen to solve the problem (41). For each bar, the design parameter, see Eq. (1), takes the initial value of 0.35, an upper bound of 1.1 is chosen as considered in [36], and a lower bound is set to 0.02. The box is set to $[-0.05, 0.05]$ for all bars, see Eq. (20). The material volume $V^{\text{target}}$ is equal to the initial value $V_0$ of the initial design. This constraint is accounted for using a penalty of 100. $N_{\text{steps}}$ and $N_{\text{failed}}$ are set to 200 and 10, respectively. The iterative process stops when the objective function no longer decreases. The evolutions of the compliance $C$ and of the ratio $\frac{V}{V^{\text{target}}}$ during the optimisation process are reported in Figure 3a in which the tests are repeated three times. It is shown that the objective function gradually decreases while satisfying the volume constraint with a relative error of the optimised solution smaller than 0.3%. The optimised configuration reported in Figure 3b is similar with the ones obtained with SA-SP reported in Figure 3c and with MMA reported ins Figure 3d.

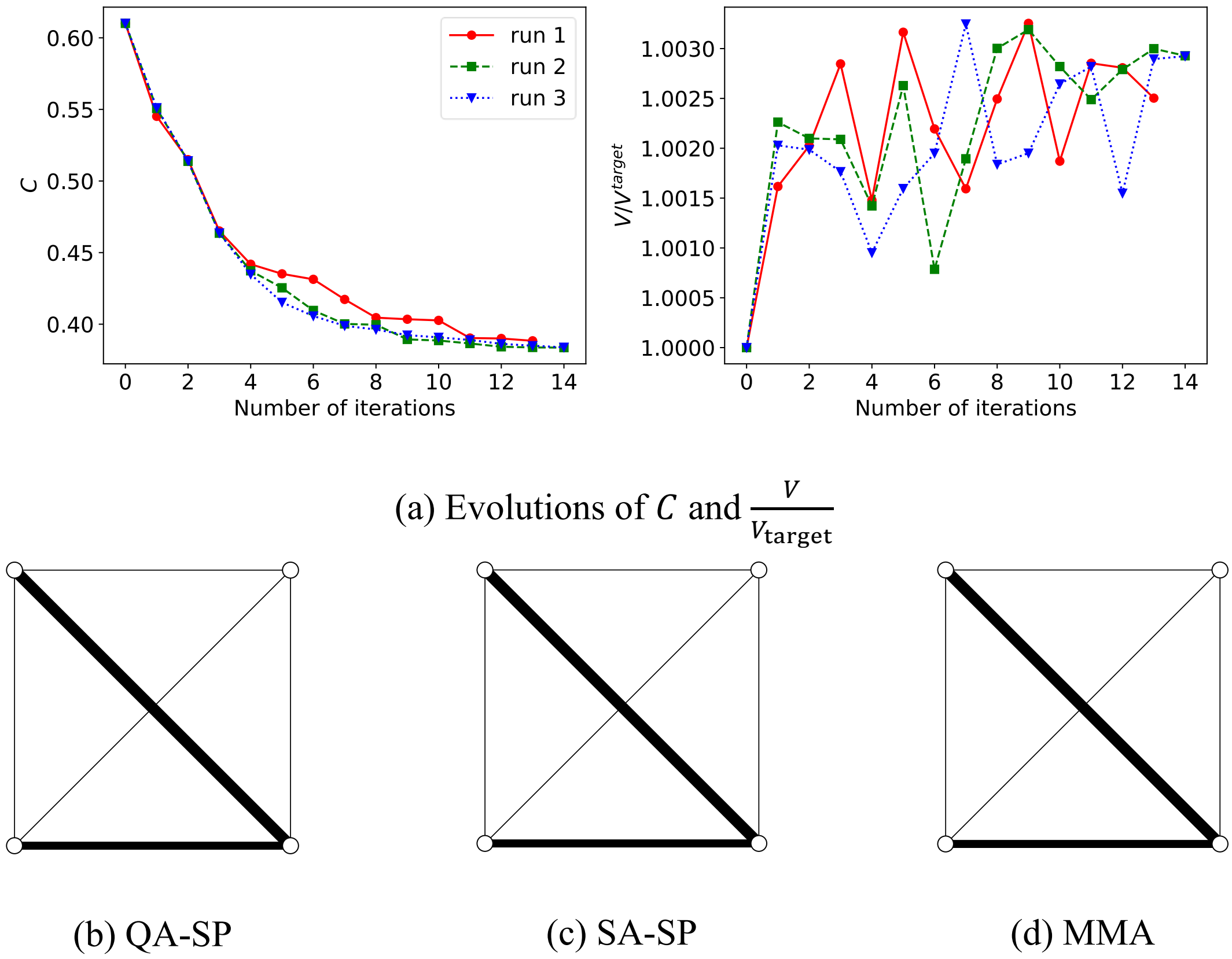

(a) Evolutions of $C$ and $\frac{V}{V_{\text{target}}}$

(b) QA-SP (c) SA-SP (d) MMA

Figure 3 - Case 1 with 6 bars: (a) evolutions of $C$ and $\frac{V}{V_{target}}$ for three independent runs and (b, c, d) optimised configurations obtained with QA-SP (the best outcome among these three independent runs), SA-SP and MMA, respectively.

The results of case 2 with 21 bars and case 3 with 29 bars are shown in Figure 4 and Figure 5 respectively, with the same parameter setting, except the initial value of the design parameter which is equal to 0.5 for each bar in the former and 0.4 in the latter. It is observed that the objective function gradually decreases in both cases. The volume constraint is satisfied with a relative error of the optimised solution smaller than 0.5% for the case of 21 bars and smaller than 0.8% for the case of 29 bars. The optimised configurations reported in Figure 4b and Figure 5b respectively for the case 2 with 21 bars and the case 3 with 29 bars are similar with the ones obtained with SA-SP reported in Figure 4c and Figure 5c and with MMA reported in Figure 4d and Figure 5d, respectively.

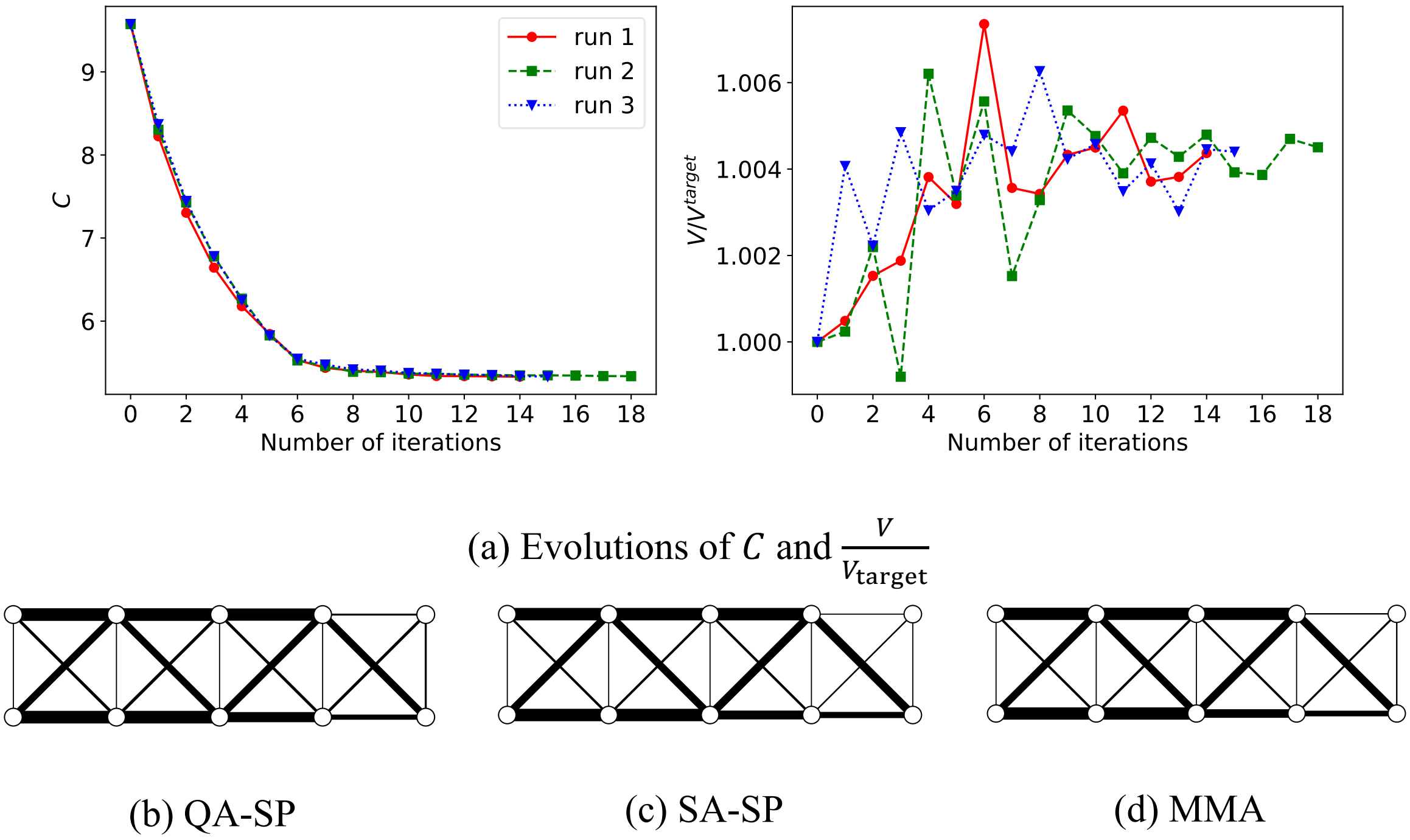

(a) Evolutions of $C$ and $\frac{V}{V_{\text{target}}}$

(b) QA-SP (c) SA-SP (d) MMA

Figure 4 - Case 2 with 21 bars: (a) evolutions of $C$ and $\frac{V}{V_{target}}$ for three independent runs and (b, c, d) optimised configurations obtained with QA-SP (the best outcome among these three independent runs), SA-SP and MMA, respectively.

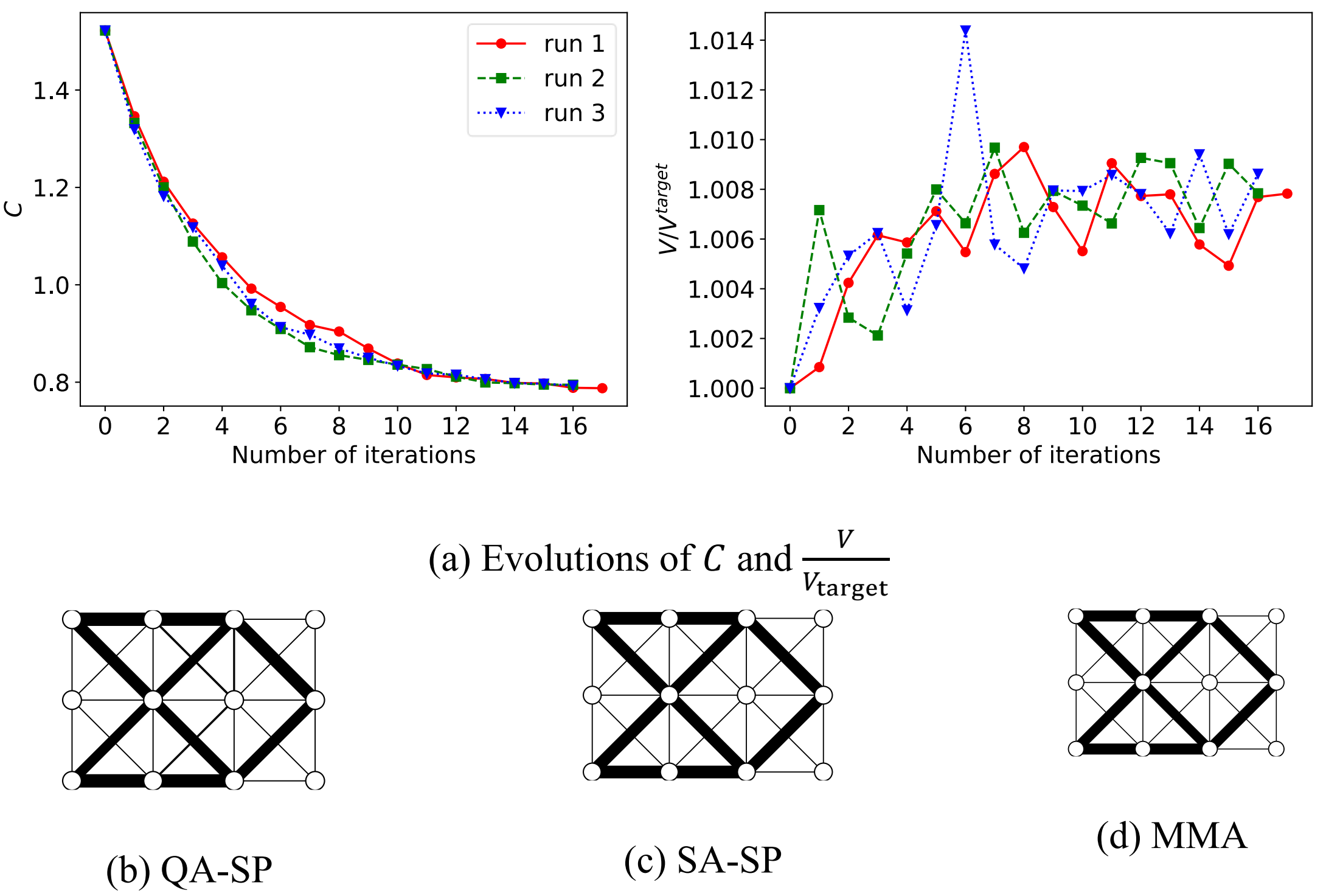

(a) Evolutions of $C$ and $\frac{V}{V_{\text{target}}}$

(b) QA-SP (c) SA-SP (d) MMA

Figure 5 - Case 3 with 29 bars: (a) evolutions of $C$ and $\frac{V}{V_{target}}$ for three independent runs and (b, c, d) optimised configurations obtained with QA-SP (the best outcome among these three independent runs), SA-SP and MMA, respectively.

We summarised in Table 1 the best objective function value reached with QA-SP at the last iteration among the three runs of the three cases reported in Figure 3, Figure 4, and Figure 5. The results obtained with SA-SP and MMA are also reported for comparison purpose. All results given by QA-SP, SA-SP and MMA are very close. The cross-sectional area of each bar in the optimised configurations for all cases are reported in Appendix A.

Table 1 - Compliance value of the optimised solution of the 6-, 21- and 29-bar cases. The results obtained with SA-SP and MMA are also reported for comparison purpose.

| Case | $K^{\mathrm{ME}}$ | $K^{\mathrm{DS}}$ | $C_{\mathrm{optimised}}$ QA-SP | $C_{\mathrm{optimised}}$ SA-SP | $C_{\mathrm{optimised}}$ MMA |
|---|---|---|---|---|---|
| 6 bars | 8 | 12 | 0.384 | 0.387 | 0.385 |
| 21 bars | 32 | 42 | 5.330 | 5.323 | 5.330 |
| 29 bars | 36 | 58 | 0.788 | 0.768 | 0.771 |

### *5.2 Three-dimensional case*

In this section, we consider a three-dimensional truss structure whose initial configuration and boundary conditions are shown in Figure 6. The initial cross-sectional area is set to 0.1 $\mathrm{m}^2$ for all bars. The parameter settings are the same as in the previous section, except the initial value which is equal to 0.1 for each bar and $N_{\mathrm{steps}}$ and $N_{\mathrm{failed}}$ are set to 300 and 20, respectively. The evolutions of the compliance $C$ and of the ratio $\frac{V}{V^{\mathrm{target}}}$ are reported in Figure 7a. It is shown that the objective function gradually decreases while satisfying the volume constraint with a relative error of the optimised solution smaller than 0.4%. The optimised configuration shown in Figure 7b is similar with the ones obtained with SA-SP reported in Figure 7c and with MMA reported ins Figure 7d. The cross-sectional area of each bar in the optimised configuration is reported in Appendix A.

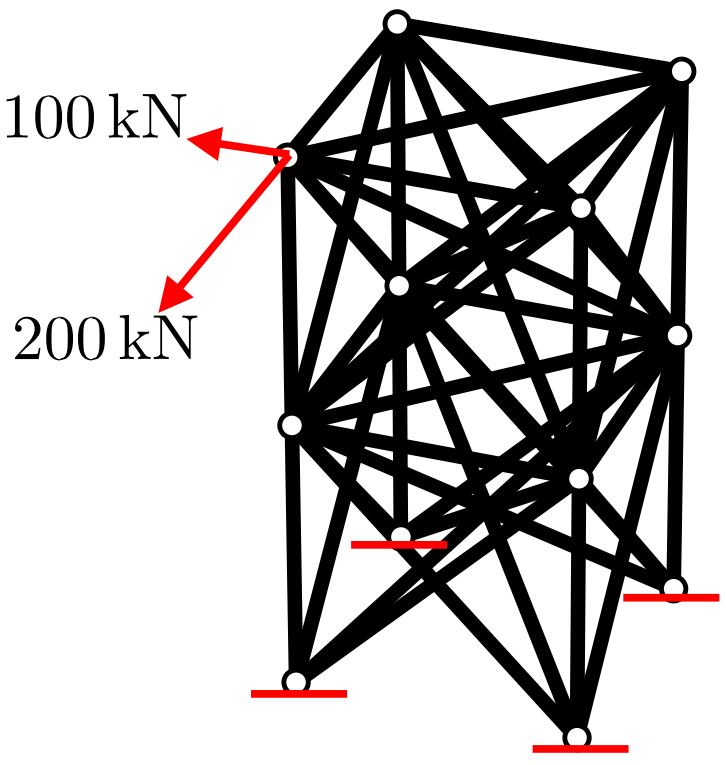

Figure 6 - Three-dimensional case consisting of 44 members. The length is set to 1 m for the horizontal and the vertical bars. The initial cross-sectional area is set to 0.1 $m^2$ for all bars.

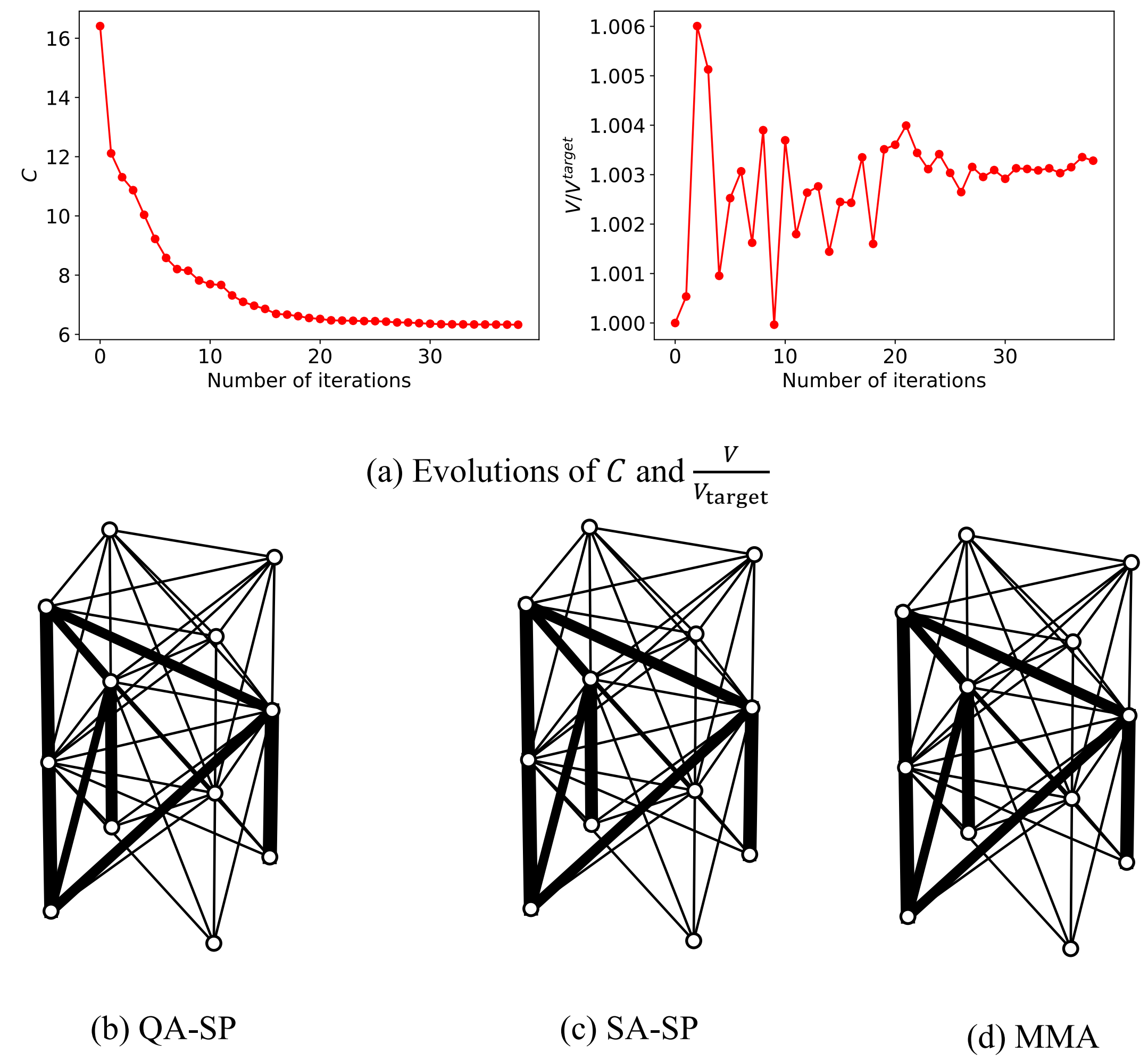

(a) Evolutions of $C$ and $\frac{V}{V_{\text{target}}}$

(b) QA-SP (c) SA-SP (d) MMA

Figure 7 - Three-dimensional case: (a) evolutions of C and $\frac{V}{V_{target}}$ and (b, c, d) optimised configurations obtained with QA-SP, SA-SP and MMA, respectively.

We summarised in Table 2 the objective function value reached with the QA-SP at the last iteration. The results obtained with SA-SP and MMA are also reported for comparison purpose. All results given by QA-SP, SA-SP and MMA are very close.

Table 2 - Compliance value of the optimised solution of the three-dimensional case. The results obtained with SA-SP and MMA are shown for comparison purpose.

| Case | $K^{\text{ME}}$ | $K^{\text{DS}}$ | $C_{\text{optimised}}$ QA-SP | $C_{\text{optimised}}$ SA-SP | $C_{\text{optimised}}$ MMA |
|---|---|---|---|---|---|
| 3D, 44 bars | 48 | 88 | 0.788 | 0.768 | 0.771 |

### *5.3 Performance discussion*

In this section, we compare the performance of the proposed QA-based hybrid framework with the simulated annealing (SA)-based counterpart. A key factor that makes the QA-based framework more efficient than the SA-based is that the time required to solve the QUBO problem using the former does not exhibit the same trend with respect to problem size (*i.e.* the number of qubits) compared to the latter. To solve the QUBO problem (see Eq. (15)) with QA, the quantum processing unit (QPU) access time can be broken down into two parts: the first part consisting of the pre- and post-processing time, which depends on the technology and the second part required for the sampling time required to repeat the annealing process. The first part of the process could be improved in the future since the field of quantum computing is rapidly evolving, and research is ongoing to improve the stability and scalability of quantum hardware. The second part, the sampling time, can be estimated as $Rt^{\max}$ , where R is the number of repeated samplings, and $t^{\max}$ is the annealing time, as defined in Eq. (11). In the tests previously reported, we set $R = 200$ and $t^{\max} = 20$ μs, as specified in Algorithm 4. In contrast, the SA-based framework requires evaluating the QUBO function with $O(RK^2)$ operations where $K$ is the number of binary variables and $R = 200$ as specified in Algorithm 5. As a result, the superior scaling of computation time with respect to $K$ in the QA-based framework highlights one of the key advantages of quantum computing over classical methods.

As illustrated in Algorithm 3, each optimization step consists of minimising the mechanical problem and updating the design. To facilitate comparison between the QA-based and SA-based frameworks, we define the following quantities:

- $T_{\mathrm{QA}}^{\mathrm{ME}}$ and $T_{\mathrm{QA}}^{\mathrm{DS}}$represent the average annealing times spent at each optimisation step to solve the mechanical problem (abbreviated by ME) and perform the design update (abbreviated by DS), respectively. These quantities are estimated as:
$$T_{\mathrm{QA}}^{\mathrm{ME}} = I^{\mathrm{ME}} R t^{\max} \text{ and } T_{\mathrm{QA}}^{\mathrm{DS}} = I^{\mathrm{DS}} R t^{\max}, \tag{48}$$
where $I^{\mathrm{ME}}$ and $I^{\mathrm{DS}}$ represent the average number of iterations needed to achieve convergence for ME and DS problems at each optimisation step, respectively.
- $T_{\mathrm{SA}}^{\mathrm{ME}}$ and $T_{\mathrm{SA}}^{\mathrm{DS}}$represent the average times spent to solve the QUBO problems at each optimisation step to solve the mechanical problem and perform the design update using SA, respectively.

Table 3 presents the values of TQAME, TSAME, TQADS, and TSADS for all performed cases in the previous sections. As expected, the QA-based approach significantly outperforms the SA-based approach in terms of time efficiency.

Table 3 - Average time to find the QUBO solution in each optimisation step of all cases

| Case | $K^{\mathrm{ME}}$ | $K^{\mathrm{DS}}$ | $T_{\mathrm{QA}}^{\mathrm{ME}}$ (s) | $T_{\mathrm{QA}}^{\mathrm{DS}}$ (s) | $T_{\mathrm{SA}}^{\mathrm{ME}}$ (s) | $T_{\mathrm{SA}}^{\mathrm{DS}}$ (s) |
|---|---|---|---|---|---|---|
| 2D, 6 bars | 8 | 12 | 0.143 | 0.04 | 0.81 | 0.253 |
| 2D, 21 bars | 32 | 42 | 1.438 | 0.04 | 4.261 | 1.371 |
| 2D, 29 bars | 36 | 58 | 0.412 | 0.04 | 4.483 | 2.244 |
| 3D, 44 bars | 48 | 88 | 0.909 | 0.08 | 3.678 | 7.147 |

### *5.4 Material nonlinearity*

In the previous sections, we focused solely on linear elastic material behaviour. In this section, we demonstrate the capability of our approach in addressing mechanical problems involving nonlinearities, such as nonlinear materials. For this purpose, we employ the nonlinear elastic behaviour governed by the following unidimensional and incompressible Mooney-Rivlin elastic potential

$$W(\lambda) = \frac{E}{6}\left(\lambda^2 + \frac{2}{\lambda} - 3\right), \tag{49}$$

where $\lambda$ is the axial stretch ratio, which is related to the strain $\varepsilon$ by $\lambda = 1 + \varepsilon$, and $E$ is a material constant. The elastic potential (49) results in the following stress-strain relationship

$$\sigma(\lambda) = \frac{dW}{d\lambda} = \frac{E}{3}\left(\lambda - \frac{1}{\lambda^2}\right). \tag{50}$$

The nonlinear stress-strain behaviour (50) is shown in Figure 8. The linearised elastic response, *i.e.* $\sigma = E(\lambda - 1)$, is also included for comparison purpose.

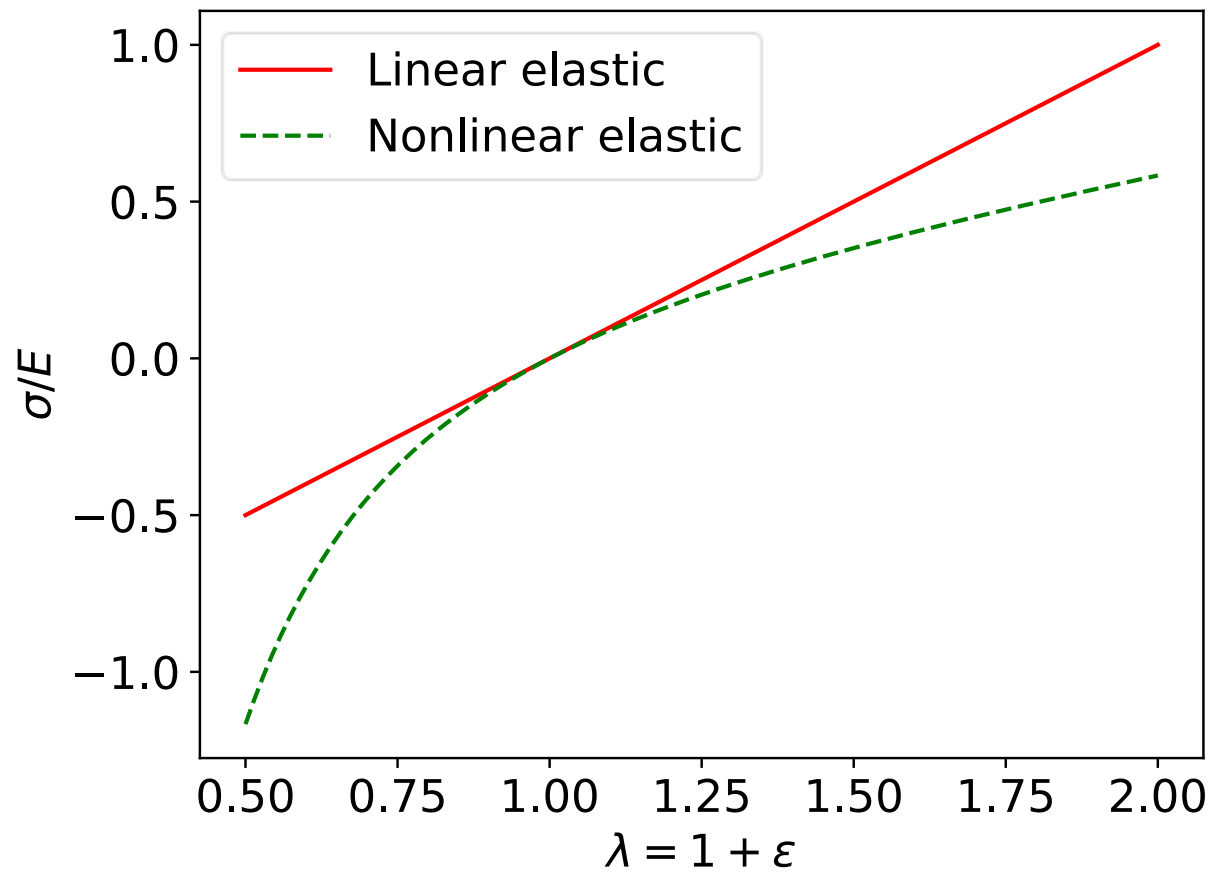

Figure 8 - Nonlinear stress-strain behaviour considered in this section. The linear elastic behaviour with the same Young's modulus is also reported for comparison purpose.

With the nonlinear elastic material governed by the elastic potential (49), the minimisation problem (7) of the mechanical part is reformulated as follows

$$\boldsymbol{U}(\boldsymbol{\alpha}) = \arg\min_{\boldsymbol{U}'} \Psi^{\mathrm{nl}}(\boldsymbol{U}'; \boldsymbol{\alpha}), \tag{51}$$

where $\Psi^{\mathrm{nl}} = \mathrm{W}^{\mathrm{int}} - \boldsymbol{F}^T\boldsymbol{U}$ represents the total potential energy with $W^{\mathrm{int}}$ being the internal energy, which is estimated by

$$\mathrm{W}^{int} = \sum_{k=1}^{N} \alpha_k \, A_k^0 L_k W(\lambda_k). \tag{52}$$

In this equation, $\lambda_k$ is the axial stretch ratio of the bar $k$ estimated as

$$\lambda_k = 1 + \varepsilon_k = 1 + \left(\frac{\boldsymbol{U}_k^{(2)} - \boldsymbol{U}_k^{(1)}}{L_k}\right) \cdot \left(\boldsymbol{x}_k^{(2)} - \boldsymbol{x}_k^{(1)}\right), \tag{53}$$

where the indices (1) and (2) denote two extremities of the bar, and $\boldsymbol{U}_k^{(i)}$ and $\boldsymbol{x}_k^{(i)}$ denote respectively the nodal displacement and the coordinates of the node $i$ of the bar $k$.

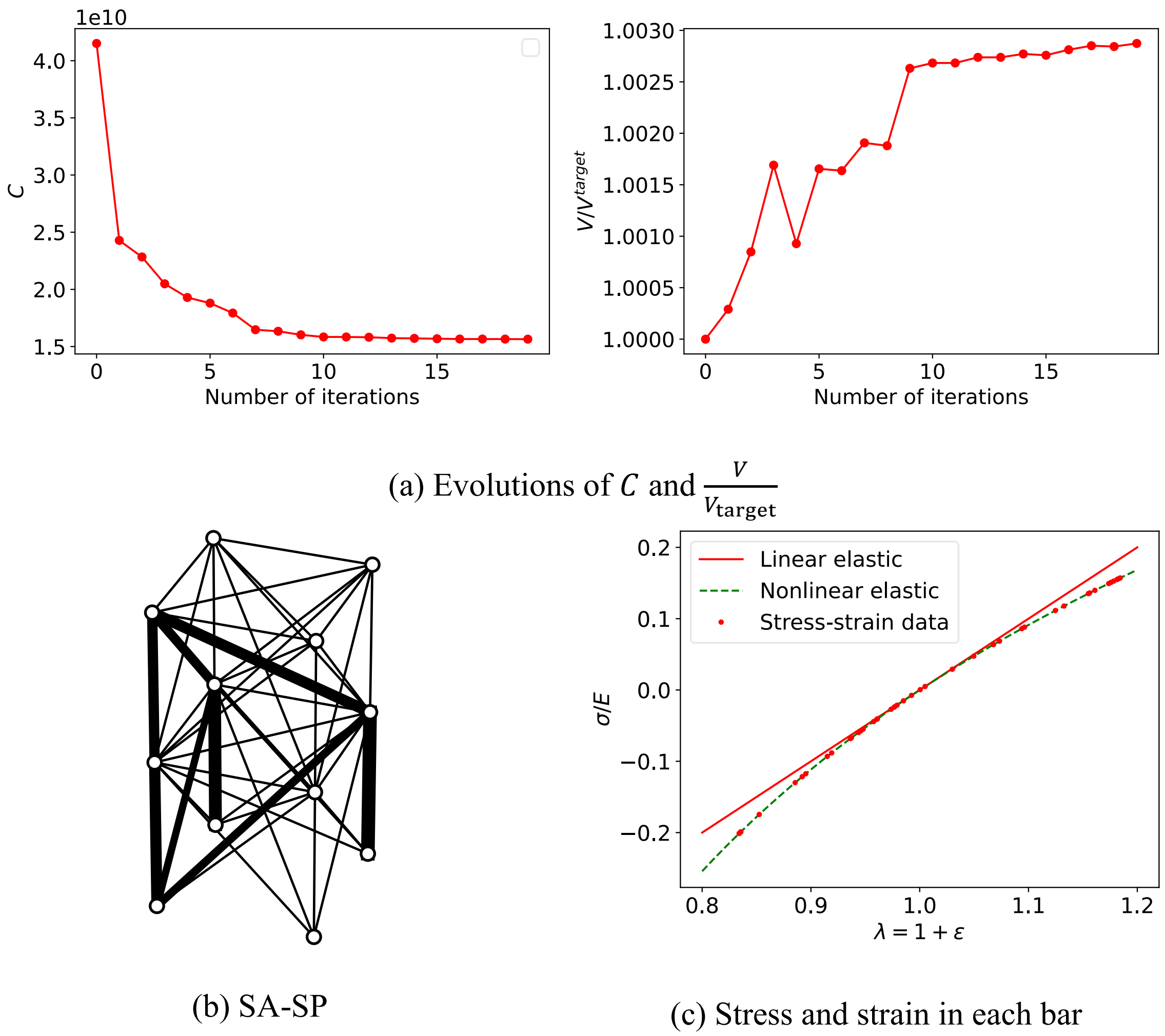

(a) Evolutions of $C$ and $\frac{V}{V_{\text{target}}}$

(b) SA-SP

(c) Stress and strain in each bar

Figure 9 - Three-dimensional case with nonlinear material behaviour: (a) evolutions of $C$ and $\frac{V}{V_{target}}$, (b) optimised configuration obtained with SA-SP, and (c) the stress and strain values in each bar.

We reconsider the three-dimensional truss structure from Section 5.2 with $E = 200$ GPa and we solve the optimisation problem with SA-SP. The applied forces are increased by a factor of 50000 to push the stress-strain behaviour into the nonlinear regime. The parameter settings remain the same as considered in Section 5.2, except for the initial discretization error $\boldsymbol{\epsilon}_0 = \frac{10^{-2}}{2^{L}-1}[1 \ldots 1]$, where $L$ is the number of qubits per continuous unknown, which is adapted since the range of displacement unknowns increases with the new applied forces. The evolutions of the objective function $C$ and of the ratio $\frac{\mathrm{V}}{\mathrm{V}^{\text{target}}}$ are reported in Figure 9(a). It is shown that the objective function gradually decreases while satisfying the volume constraint with a relative error of the optimised solution smaller than 0.3%. The optimised configuration shown in Figure 9(b) is similar with the ones obtained in Section 5.2. The stress-strain values in each bar are reported in Figure 9(c), in which the strain can reach 0.2 in the nonlinear regime. The cross-sectional area of each bar in the optimised configuration is reported in Appendix A.

## 6. Conclusions

In this study, we proposed a novel hybrid framework that combines classical computers and quantum annealers for optimising truss structures, leveraging the strength of quantum annealing to solve both the mechanical boundary value problem and the constrained minimisation problem required to adjust the design variables. Unlike classical optimisation methods, our hybrid framework harnesses the quantum annealing technique to find the better configuration after each optimisation step without the needs of performing arithmetic operations. This innovative approach has shown promising potential to enhance the optimisation process, particularly in handling complex, nonlinear, and high-dimensional solution spaces that are common in structural design.

However, the hybrid framework faces challenges related to the limited availability and scalability of current quantum technologies. As quantum hardware continues to advance, we anticipate that these challenges will be addressed, further boosting the applicability of the framework to real-world engineering problems.

The proposed framework can be extended beyond truss structure optimisation. In future work, the proposed framework will be adapted to carry out continuous topology optimisation and to integrate a catalogue of materials and unit cells to be distributed in the design space to form mixed discrete-continuous optimisation problems. The higher-order Taylor's series will be considered to improve the approximation accuracy of the objective functions and constraints in the framework, leading to optimising higher-order polynomial forms, which requires solving higher-order unconstrained binary optimisation (HUBO) formulations after applying the continuous-binary transformation. These directions open promising possibilities for applying the hybrid approach to a wide range of engineering and computational design problems with linear and non-linear behaviours.

## Acknowledgments

This work is supported by the European Regional Development Fund (ERDF/FEDER) and the Walloon Region of Belgium through project 925 VirtualLab_Cenaero (programme 2021-2027). This work is partially supported by a ''Strategic Opportunity'' grant from the University of Liege.

## Appendix A. Optimised solutions

The cross-sectional area of each bar in the optimised configuration is reported in Table 4, Table 5, and Table 6 for the three two-dimensional cases in Section 5.1 and Table 7 for the three dimensional case in Section 5.2, respectively.

Table 4 - Optimised solution: cross-sectional area of each bar for Case 1 - 6 bars. Each bar is identified by the coordinates of its extremities.

| Index | Bar | Area ratio |
|---|---|---|
| 1 | (0,0)-(1,0) | 0.384 |
| 2 | (1,0)-(1,1) | 0.01 |
| 3 | (1,1)-(0,1) | 0.01 |
| 4 | (0,0)-(1,1) | 0.01 |
| 5 | (1,0)-(0,1) | 0.545 |
| 6 | (0,0)-(0,1) | 0.01 |

Table 5 - Optimised solution: cross-sectional area of each bar for Case 2 - 21 bars. Each bar is identified by the coordinates of its extremities.

| Index | Bar | Area ratio | Index | Bar | Area ratio |
|---|---|---|---|---|---|
| 1 | (0,0)-(1,0) | 0.55 | 12 | (2,0)-(3,0) | 0.442 |
| 2 | (1,0)-(1,1) | 0.01 | 13 | (3,0)-(3,1) | 0.0104 |
| 3 | (1,1)-(0,1) | 0.55 | 14 | (3,1)-(2,1) | 0.533 |
| 4 | (0,0)-(1,1) | 0.312 | 15 | (2,0)-(3,1) | 0.315 |
| 5 | (1,0)-(0,1) | 0.131 | 16 | (3,0)-(2,1) | 0.124 |
| 6 | (0,0)-(0,1) | 0.01 | 17 | (3,0)-(4,0) | 0.225 |
| 7 | (1,0)-(2,0) | 0.55 | 18 | (4,0)-(4,1) | 0.0881 |
| 8 | (2,0)-(2,1) | 0.0106 | 19 | (4,1)-(3,1) | 0.0824 |
| 9 | (2,1)-(1,1) | 0.55 | 20 | (3,0)-(4,1) | 0.126 |
| 10 | (1,0)-(2,1) | 0.136 | 21 | (4,0)-(3,1) | 0.314 |
| 11 | (2,0)-(1,1) | 0.305 | | | |

Table 6 - Optimised solution: cross-sectional area of each bar for Case 3 - 29 bars. Each bar is identified by the coordinates of its extremities.

| Index | Bar | Area ratio | Index | Bar | Area ratio |
|---|---|---|---|---|---|
| 1 | (0,0)-(1,0) | 0.55 | 16 | (3,0)-(2,1) | 0.0398 |
| 2 | (1,0)-(1,1) | 0.0401 | 17 | (1,1)-(1,2) | 0.01 |
| 3 | (1,1)-(0,1) | 0.0113 | 18 | (1,2)-(0,2) | 0.549 |
| 4 | (0,0)-(1,1) | 0.41 | 19 | (0,1)-(1,2) | 0.0487 |
| 5 | (1,0)-(0,1) | 0.0218 | 20 | (1,1)-(0,2) | 0.493 |
| 6 | (0,0)-(0,1) | 0.0161 | 21 | (0,1)-(0,2) | 0.0104 |
| 7 | (1,0)-(2,0) | 0.544 | 22 | (2,1)-(2,2) | 0.0811 |
| 8 | (2,0)-(2,1) | 0.0408 | 23 | (2,2)-(1,2) | 0.54 |
| 9 | (2,1)-(1,1) | 0.01 | 24 | (1,1)-(2,2) | 0.401 |
| 10 | (1,0)-(2,1) | 0.0877 | 25 | (2,1)-(1,2) | 0.0908 |
| 11 | (2,0)-(1,1) | 0.465 | 26 | (3,1)-(3,2) | 0.0239 |
| 12 | (2,0)-(3,0) | 0.0108 | 27 | (3,2)-(2,2) | 0.0221 |
| 13 | (3,0)-(3,1) | 0.0261 | 28 | (2,1)-(3,2) | 0.0103 |
| 14 | (3,1)-(2,1) | 0.0396 | 29 | (3,1)-(2,2) | 0.528 |
| 15 | (2,0)-(3,1) | 0.459 | | | |

Table 7 - Optimised solution: cross-sectional area of each bar for the linear three-dimensional case - 44 bars. Each bar is identified by the coordinates of its extremities.

| Index | Bar | Area ratio | Index | Bar | Area ratio |
|---|---|---|---|---|---|
| 1 | (0,0,0)-(0,0,1) | 0.288 | 23 | (0,0,1)-(0,0,2) | 0.275 |
| 2 | (0,0,0)-(1,0,1) | 0.0103 | 24 | (0,0,1)-(1,0,2) | 0.0102 |
| 3 | (0,0,0)-(1,1,1) | 0.219 | 25 | (0,0,1)-(1,1,2) | 0.0103 |
| 4 | (0,0,0)-(0,1,1) | 0.177 | 26 | (0,0,1)-(0,1,2) | 0.0147 |
| 5 | (1,0,0)-(0,0,1) | 0.0106 | 27 | (1,0,1)-(0,0,2) | 0.0104 |
| 6 | (1,0,0)-(1,0,1) | 0.0134 | 28 | (1,0,1)-(1,0,2) | 0.0103 |
| 7 | (1,0,0)-(1,1,1) | 0.0101 | 29 | (1,0,1)-(1,1,2) | 0.0103 |
| 8 | (1,0,0)-(0,1,1) | 0.01 | 30 | (1,0,1)-(0,1,2) | 0.0103 |
| 9 | (1,1,0)-(0,0,1) | 0.011 | 31 | (1,1,1)-(0,0,2) | 0.211 |
| 10 | (1,1,0)-(1,0,1) | 0.0112 | 32 | (1,1,1)-(1,0,2) | 0.01 |
| 11 | (1,1,0)-(1,1,1) | 0.299 | 33 | (1,1,1)-(1,1,2) | 0.0126 |
| 12 | (1,1,0)-(0,1,1) | 0.0195 | 34 | (1,1,1)-(0,1,2) | 0.0106 |
| 13 | (0,1,0)-(0,0,1) | 0.0172 | 35 | (0,1,1)-(0,0,2) | 0.189 |
| 14 | (0,1,0)-(1,0,1) | 0.01 | 36 | (0,1,1)-(1,0,2) | 0.0102 |
| 15 | (0,1,0)-(1,1,1) | 0.01 | 37 | (0,1,1)-(1,1,2) | 0.01 |
| 16 | (0,1,0)-(0,1,1) | 0.251 | 38 | (0,1,1)-(0,1,2) | 0.0113 |
| 17 | (0,0,1)-(1,0,1) | 0.0108 | 39 | (0,0,2)-(1,0,2) | 0.01 |
| 18 | (0,0,1)-(1,1,1) | 0.01 | 40 | (0,0,2)-(1,1,2) | 0.0143 |
| 19 | (0,0,1)-(0,1,1) | 0.0103 | 41 | (0,0,2)-(0,1,2) | 0.0159 |
| 20 | (1,0,1)-(1,1,1) | 0.0105 | 42 | (1,0,2)-(1,1,2) | 0.0104 |
| 21 | (1,0,1)-(0,1,1) | 0.0103 | 43 | (1,0,2)-(0,1,2) | 0.0105 |
| 22 | (1,1,1)-(0,1,1) | 0.0113 | 44 | (1,1,2)-(0,1,2) | 0.0101 |

Table 8 - Optimised solution: cross-sectional area of each bar for the nonlinear three-dimensional case - 44 bars. Each bar is identified by the coordinates of its extremities.

| Index | Bar | Area ratio | Index | Bar | Area ratio |
|---|---|---|---|---|---|
| 1 | (0,0,0)-(0,0,1) | 0. 256 | 23 | (0,0,1)-(0,0,2) | 0.237 |
| 2 | (0,0,0)-(1,0,1) | 0. 01 | 24 | (0,0,1)-(1,0,2) | 0.01 |
| 3 | (0,0,0)-(1,1,1) | 0.19 | 25 | (0,0,1)-(1,1,2) | 0.01 |
| 4 | (0,0,0)-(0,1,1) | 0. 166 | 26 | (0,0,1)-(0,1,2) | 0.01 |
| 5 | (1,0,0)-(0,0,1) | 0.01 | 27 | (1,0,1)-(0,0,2) | 0.01 |
| 6 | (1,0,0)-(1,0,1) | 0.01 | 28 | (1,0,1)-(1,0,2) | 0.01 |
| 7 | (1,0,0)-(1,1,1) | 0.01 | 29 | (1,0,1)-(1,1,2) | 0.01 |
| 8 | (1,0,0)-(0,1,1) | 0.01 | 30 | (1,0,1)-(0,1,2) | 0.01 |
| 9 | (1,1,0)-(0,0,1) | 0.01 | 31 | (1,1,1)-(0,0,2) | 0.256 |
| 10 | (1,1,0)-(1,0,1) | 0.01 | 32 | (1,1,1)-(1,0,2) | 0.01 |
| 11 | (1,1,0)-(1,1,1) | 0.313 | 33 | (1,1,1)-(1,1,2) | 0.01 |
| 12 | (1,1,0)-(0,1,1) | 0.0204 | 34 | (1,1,1)-(0,1,2) | 0.01 |
| 13 | (0,1,0)-(0,0,1) | 0.01 | 35 | (0,1,1)-(0,0,2) | 0.22 |
| 14 | (0,1,0)-(1,0,1) | 0.01 | 36 | (0,1,1)-(1,0,2) | 0.01 |
| 15 | (0,1,0)-(1,1,1) | 0.01 | 37 | (0,1,1)-(1,1,2) | 0.01 |
| 16 | (0,1,0)-(0,1,1) | 0.294 | 38 | (0,1,1)-(0,1,2) | 0.01 |
| 17 | (0,0,1)-(1,0,1) | 0.01 | 39 | (0,0,2)-(1,0,2) | 0.01 |
| 18 | (0,0,1)-(1,1,1) | 0.01 | 40 | (0,0,2)-(1,1,2) | 0.01 |
| 19 | (0,0,1)-(0,1,1) | 0.01 | 41 | (0,0,2)-(0,1,2) | 0.0144 |
| 20 | (1,0,1)-(1,1,1) | 0.01 | 42 | (1,0,2)-(1,1,2) | 0.01 |
| 21 | (1,0,1)-(0,1,1) | 0.01 | 43 | (1,0,2)-(0,1,2) | 0.01 |
| 22 | (1,1,1)-(0,1,1) | 0.01 | 44 | (1,1,2)-(0,1,2) | 0.01 |